\documentstyle[11pt]{article}
\newcommand {\wop}{\ \omega^{\prime}_0 \ }
\newcommand {\wopp}{\ \omega^{\prime \prime}_0 \ }

\newcommand {\wunp}{\ \omega^{\prime}_1 \ }
\newcommand {\wunpp}{\ \omega^{\prime \prime}_1 \ }

\newcommand {\wdos}{\ \omega_2}
\newcommand {\wdosp}{\ \omega^{\prime}_2 \ }
\newcommand {\wdospp}{\ \omega^{\prime \prime}_2 \ }

\newcommand {\wtres}{\ \omega_3}
\newcommand {\wtresp}{\ \omega^{\prime}_3 \ }
\newcommand {\wtrespp}{\ \omega^{\prime \prime}_3 \ }

\newcommand {\wcinc}{\ \omega_5 }
\newcommand {\wcincp}{\ \omega^{\prime}_5 \ }
\newcommand {\wcincpp}{\ \omega^{\prime \prime}_5 \ }

\newcommand {\wsisp}{\ \omega^{\prime}_6 \  }
\newcommand {\wsispp}{\ \omega^{\prime \prime}_6 \  }

\newcommand {\nz}{\ n_l}
\newcommand {\vzd}{\ v_{02}  }

\newcommand {\vtu}{\ v_{31}  }

\begin{document}
\newcommand{\ECM}{\em Departament d'Estructura i Constituents de la
Mat\`eria
                  \\ Facultat de F\'\i sica, Universitat de Barcelona \\
                     Diagonal 647, E-08028 Barcelona, Spain \\
                                and \\
                         I.F.A.E.}

\def\thefootnote{\fnsymbol{footnote}}
\pagestyle{empty}
{\hfill \parbox{6cm}{\begin{center} UB-ECM-PF 96/16\\
                                    hep-ph/9610549\\
                                    October 1996
                     \end{center}}}
\vspace{1.5cm}

\begin{center}
\large{Chiral Effective Lagrangian in the large-$N_c$ limit: 
the nonet case.}

\vskip .6truein
\centerline {P. Herrera-Sikl\'ody\footnote{e-mail: herrera@sophia.ecm.ub.es},
J.I. Latorre\footnote{e-mail: latorre@sophia.ecm.ub.es},
P. Pascual\footnote{e-mail: pascual@ababel.ecm.ub.es}
and J. Taron\footnote{e-mail: taron@ifae1.ecm.ub.es}}
\end{center}
\vspace{.3cm}
\begin{center}
\ECM
\end{center}
\vspace{1.5cm}

\centerline{\bf Abstract}
\medskip
A $U_L(3) \otimes U_R(3)$
low-energy effective lagrangian for the nonet of pseudogoldstone bosons
that appear in the large $N_c$ limit of QCD
is presented including terms 
up to four derivatives and explicit symmetry breaking terms up to 
quadratic in the quark masses.
The one-loop renormalization of the couplings is worked 
out using the heat-kernel technique and dimensional renormalization.
The calculation is carried through for $U_L(n_l)\otimes U_R(n_l)$, thus
allowing for a generic number $n_l$ of light quark flavours. The
crucial advantages that the expansion in powers of $1/N_c$ bring
about are discussed. Special emphasis is put in pointing out what features
are at variance with the $SU_L\otimes SU_R$ results when the singlet
$\eta'$ is included in the theory.

\bigskip



\newpage
\pagestyle{plain}

\section{The $U_L(3) \otimes U_R(3)$ symmetry.}

The pattern of the lowest-lying states
in the spectrum of strong interactions uncovers an approximate continuous
symmetry of nature, the so-called {\it chiral symmetry}, which is 
spontaneously broken. The octet of pseudoscalar particles
- $\pi$, $K$, and $\eta$ -, with
masses much smaller than those of the next excited states - the octet
of vector particles $\rho$, $\omega$ and $K^*$, the baryons-, are the
accepted candidates for pseudo-goldstone bosons associated to the spontaneous
breaking of the symmetry. 

This approximate symmetry is well incorporated in QCD as three of
the quarks happen to be light. In the (chiral) limit of vanishing 
$m_u,\; m_d, \; m_s$ the QCD lagrangian has the symmetry freedom of
arbitrarily
rotating with unitary matrices the quark field components in the space
of flavours (u,d,s), independently in the Left and the Right sectors of 
chirality eigenstates. The symmetry group is
$U_L(3) \otimes U_R(3)$ and is explicitly broken
by the light quark masses; if it had not,
the lightest mesons would indeed have been massless particles.
This breaking is small, though,
since the light quark masses are much smaller than the typical hadronic 
scale of a few hundred MeV. (This is certainly so for the $u$ and $d$
quarks ( $2 < m_u < 8$ and $5 < m_d < 15$ MeV) and still approximately verified
for the heavier $s$-quark ( $100 < m_s < 300$ MeV), 
\cite{pdb96}). 

Empirically, however, only the $SU_L(3) \otimes SU_R(3)$ symmetry subgroup,
spontaneously broken to $SU_{L+R}(3)$, 
is manifest: instead
of a nonet of light pseudoscalars only an octet is observed. Of the remaining
$U_L(1) \otimes U_R(1)$, the vector part provides the conserved baryon 
number current,
whereas the axial $U_A(1)$
does not seem reflected at all in the spectrum,
either as a conserved quantum number or as a goldstone boson.
The first possibility would imply that all massive hadrons would
appear in parity doublets and this is not what is observed.
On the other hand,
by its quantum numbers the $\eta'$ would be the ninth candidate
for goldstone boson; but if the $U_A(1)$ were realized in the Goldstone
mode and explicitly broken only by the same quark mass terms
that break the $SU_L(3) \otimes SU_R(3)$ one would expect that the
ninth pseudo-goldstone boson would have a mass similar to the pion
(to the masses in the octet): actually, a singlet pseudoscalar meson ought to
exist with a mass smaller than  $\sqrt{3} m_{\pi}$ \cite{weinberg1}. 
Such a particle is missing in the spectrum. There is the $\eta'$ instead
but it is indeed so much heavier than the
$\pi$, $K$, $\eta$ that, at any rate, it seems 
hard to conceive it on the same footing as the light pseudoscalars,
altogether in a nonet.

This puzzle is part of what is known as
the $U_A(1)$ problem of QCD \cite{sksw}, and originates
in that the $U_A(1)$ symmetry of the QCD lagrangian
is anomalous at the quantum
level. Yet a conserved ninth singlet axial current can still 
be defined in spite of the $U_A(1)$ anomaly, it is not gauge invariant and
't Hooft showed how its conservation could be broken by non-perturbative 
effects \cite{hooft}. 

Nevertheless, there is a limit of QCD in which the $\eta '$
appears as the ninth, genuine goldstone boson: the limit of large number 
of colours $N_c$.
't Hooft proposed a systematic expansion of QCD \cite{hooftNc} \cite{Witten}
with the
inverse number of colours, $1/N_c$, as the expansion parameter. With the
only assumption that in the framework of this $1/N_c$ expansion QCD confines,
a few qualitative features of the strong interactions already emerge
by keeping only the leading terms. Of special interest to us is the result
that the pattern of 
chiral symmetry breaking is exactly
$U_L(n_l) \otimes U_R(n_l) \to U_{L+R}(n_l)$, where $n_l$ stands for
a generic number of quark flavours \cite{cw}, which resembles very much the
pattern for chiral symmetry breaking observed in nature. Now
in the $N_c \to \infty$ limit there is no
spoilt $U_A(1)$ anymore, since the anomaly 
in the divergence of the singlet axial current
is $1/N_c$ suppressed and the full $U_L \otimes U_R$ is recovered
along with the entire nonet of goldstone bosons \cite{w1} \cite{v}.

The present study is based on the framework that the $1/N_c$ expansion
provides. Following the steps pioneered by  Weinberg \cite{weinberg} and
Gasser and Leutwyler \cite{gl}, we write the low-energy effective lagrangian
of the  $N_c \to \infty$ limit of QCD, with $m_u=m_d=m_s=0$: it is 
a chiral lagrangian that involves the whole nonet of pseudo-goldstone
bosons and is invariant under $U_L(3) \otimes U_R(3)$. 
The departures from this scenario, which stem from the explicit breaking
of chiral symmetry by quark masses and from the $U_A(1)$ anomaly,
are treated perturbatively, in powers of the 
quark masses and $1/N_c$. It is conceivable that a good
picture of the lightest hadrons and their interactions at 
low energies could emerge from this approach.

Many authors \cite{vrw} have discussed how to construct such lagrangian,
{\it i.e.}, how to extend the symmetry to $U_L(3) \otimes U_R(3)$ and
properly take into account the effects of the $U_A(1)$ anomaly at the
same time,
nicely
organized in powers of $1/N_c$. In these articles the physical consequences
to lowest orders in $1/N_c$ and the derivative expansion have been worked 
out as well. We closely follow their work.

The present study is devoted to extend the analysis to a full $O(p^4)$
chiral lagrangian, {\it i.e.}, with terms kept up to four derivatives
and quadratic in the quark masses. The conservative bookkeeping
of quark masses as two chiral powers, $m_q=O(p^2)$, is adopted, as in
\cite{gl} (see also \cite{gorgw}).
The celebrated Gell-Mann Okubo relations amongst the
light pseudoscalar masses squared follow from
this chiral power counting in a rather natural way.
The exact mechanism of chiral symmetry breaking being unknown,
however, many different possibilities other than $m_q=O(p^2)$ have
not been ruled out hitherto, either from the experiment or from QCD.
None of them has been considered here. They can be adequately treated
in the framework of {\it Generalized Chiral Perturbation Theory} as
proposed in \cite{fss}, if needed. In order to discriminate 
amongst the various possibilities, the nature of chiral symmetry breaking
needs to be established on more solid grounds. 

Our study is to all orders in the large-$N_c$ expansion - in a sense
that will be later qualified.

We calculate all one-loop divergences to the effective action
using the heat-kernel technique and dimensional regularisation, and
carry out to this approximation the renormalization of the couplings.

The article is organized as follows: in section 2 the method of external
sources and the generating functional are briefly reviewed and the 
notation is set. In section 3 the chiral lagrangian including terms up to
$O(p^4)$ and the one-loop effective action are put forward, before the
$1/N_c$ expansion is performed. Next, in section 4, the calculation is
organized in powers of $1/N_c$ and the first non-trivial terms are also
given. The conclusions and the appendices follow.

\section{The method of the external sources.}

In this section we briefly review the symmetries of QCD that are relevant
for the chiral lagrangian in the large-$N_c$ limit.

The QCD lagrangian can always be written with a diagonal quark mass term, as

\begin{equation}
\label{2.1}
{\cal L}_{QCD}= -\frac {1}{2}\; Tr \left(G_{\mu \nu} G^{\mu \nu} \right) 
+{\sum}^{n_l}_{f=1} 
\bar{q}_f 
\left( i \gamma_{\mu} D^{\mu} - m_f \right) q_f \;,
\end{equation}
where $f$ labels the light quarks $q_f$ 
that appear in $n_l$ number of light flavours.
In nature, $n_l=3$ at most, for the flavours $u$, $d$, $s$. 
Although explicitly omitted, the quark fields
also carry 
a colour index $q^c$ which labels the fundamental representation of
the gauge group $SU(N_c)$ of colour. Nature has chosen $N_c=3$.
The covariant derivative $D_{\mu}$
acts on colour indices through the gluon matrix, $G_{\mu}$ which
form an adjoint $(N_c^2 -1)$-dimensional representation of $SU(N_c)$.
On the quark fields it acts in the usual way, diagonal in flavour indices,
\begin{equation}
\label{2.2 }
D_{\mu} q^c = \partial_{\mu} q^c - i {\frac {g}{\sqrt{N_c}}} 
{ \left( G_{\mu} \right) }^c_{c'} \; q^{c'} \;.
\end{equation}

The field strength matrix is
$G_{\mu \nu}=\partial_\mu G_\nu -\partial_\nu G_\mu -
i {\frac {g}{\sqrt{N_c}}} \left[ G_\mu, G_\nu \right]$, whereas 
$G_{\mu} \equiv G_{\mu}^a \left( \frac {\lambda^a}{2} \right)$; 
the sum over $a$ runs from $1$ to $N_c^2 -1$
and $(\lambda^a)$ stands for the $N_c \times N_c$
Gell-Mann matrices of $SU(N_c)$.

The heavy quarks $c$, $b$, $t$ are also omitted from the lagrangian in 
(\ref{2.1}) for they decouple from the strong low energy processes which only 
involve light pseudoscalars with $u$, $d$, $s$ quantum numbers.

The $N_c$ dependence that accompanies the coupling
constant, $\frac {g}{\sqrt{N_c}}$, is explicitly displayed. 
Apart from the usual combinatorial factors that appear in the
Feynman diagrams, this extra dependence in $N_c$ must be added in 
order to allow for a smooth, non-trivial $N_c \to \infty$ limit of QCD. 
We shall revert to the issue
of counting the leading powers of $N_c$ of each effective coupling in 
section 4.

If the light quark masses are switched off, the lagrangian (\ref{2.1})
becomes invariant under the symmetry group $U_L(n_l) \otimes U_R(n_l)$,
$n_l=3$. Henceforth we keep the number of light flavours, $n_l$, generic
in the expressions.
This is most easily seen by writing the lagrangian above in terms
of the quark left and right components,
\begin{equation}
\label{2.3}
q_L= \frac {1- \gamma_5}{2} q \;,\;\;\;\;\;\;\;\;\;\;\;\;\;
q_R= \frac {1+ \gamma_5}{2} q.
\end{equation}
The terms in (\ref{2.1}) that involve the quark fields read,
\begin{equation}
\label{2.4}
\bar{q}_{L} ( i \gamma_{\mu} D^{\mu})  q_{L}+
\bar{q}_{R} ( i \gamma_{\mu} D^{\mu})  q_{R}  \;. 
\end{equation}
The symmetry of rotating independently the components $(q_L)_f$ and
$(q_R)_f$ in flavour space with unitary matrices is manifest. 
It is a global symmetry that
is explicitly broken by the quark masses. However, if all the quark
masses were the same there would still be an invariance under the
diagonal vector subgroup $U_{L+R} ({n_l})$.
In that case the subgroup coincides with the unbroken
subgroup after spontaneous breaking of $U_L(n_l) \otimes U_R (n_l)$. 

\vspace*{0.7cm}
As it is well known not all the symmetries of the classical action 
are maintained at the quantum level; 
the quantum theory thus generates anomalous contributions
to the divergences of some currents - they are no longer conserved.
 
The low-energy effective action is a convenient bookkeeping device
to encode the symmetries of the underlying theory - QCD -,
which automatically incorporates
all the unitarity features of quantum field theory.
In writing the effective lagrangian care must be taken that all the 
(chiral) Ward Identities (WI) among the Green's functions are well
implemented, including the anomalous ones.
Actually, the method put forward in ref. \cite{gl} constructs 
a solution to these WI's. It is based upon the transformation
properties of the generating functional.

The probability amplitude
of transition from the vacuum in the remote past to the vacuum in the
far future, in the presence of terms in the lagrangian that couple
the external sources linearly to the currents, as in (\ref{2.6}) below,
contains all the Green's functions among these currents.
Its logarithm $Z[f]$ is the generating functional of the connected 
Green's functions,
\begin{equation}
\label{2.5}
e^{iZ[f]}= {\sum}_n \frac {i^n}{n!} \int dx_1 dx_2  ... dx_n \;
f_{i_1}^{\mu_1}(x_1) f_{i_2}^{\mu_2}(x_2) ... f_{i_n}^{\mu_n}(x_n)
\langle 0| T \; J^{i_1}_{\mu_1}(x_1) J^{i_2}_{\mu_2}(x_2) ... 
J^{i_n}_{\mu_n}(x_n) |0\rangle ,
\end{equation}
$J'$s and $f'$s stand for generic currents and external sources, respectively.

We shall consider both
bilinear quark operators (currents) and the topological charge operator
coupled to external sources, added to the QCD lagrangian,
\begin{eqnarray}
{\cal L}&=&{\cal L}_{QCD}+ \bar{q}_L \gamma_\mu l^\mu (x) q_L 
+ \bar{q}_R \gamma_\mu r^\mu (x) q_R
- \bar{q}_R (s(x) + i p(x) ) q_L  \nonumber \\
        &-& \bar{q}_L (s(x) - i p(x) ) q_R
- \frac {g^2}{ 16 \pi^2} \frac {\theta (x)}{N_c} Tr \left( G_{\mu \nu} 
{\tilde{G}}^{\mu \nu} \right),
\label{2.6}
\end{eqnarray}
${\tilde{G}}^{\mu \nu}= \epsilon^{\mu \nu \alpha \beta} G_{\alpha \beta} $. 
The first two new terms correspond to sources for the 
$U_L \otimes U_R$ Noether currents of QCD, the non-singlets are
the generators of Current Algebra;
$s$ is a source for the quark mass term and $p$ for pseudoscalar bilinears
with the quantum numbers of $\pi$, $K$, $\eta$ and $\eta'$.
The sources
$l_\mu$, $r_\mu$, $s$ and $p$ are hermitian $n_l \times n_l$
matrices; $\theta$ is a real function. The axial $a_\mu$ and the
vector $v_\mu$ sources are defined so that $l_\mu=v_\mu - a_\mu$ and
$r_\mu=v_\mu + a_\mu$. One can, formally, write the generating functional
as a path integral,
\begin{equation}
\label{2.7}
exp \{i Z[l, r, s, p, \theta] \}= \int [d\bar{q} \; dq \; dG_\mu]  \; 
exp \{i \int dx {\cal L} \}.
\end{equation}
The connected Green's functions are obtained by performing functional 
derivatives of $Z$ with respect to the sources. 

In order to further constrain the form of the effective lagrangian
it is a convenient trick
to promote the global $U_L \otimes U_R$ transformations
- that leave the QCD lagrangian invariant - to local ones 
by allowing  the external sources to 
transform along with the dynamical fields. 
The combined set of local $U_L \otimes U_R$ 
transformations\footnote{Being unitary,
$g_R$ and $g_L$ can be always parametrized as $g_R=exp(i \beta) exp(i \alpha)$,
$g_L=exp(i \beta) exp(-i \alpha)$, with $\alpha$ and $\beta$ $n_l \times
n_l$ hermitian matrices. A pure vector transformation has $\alpha=0$,
whereas an axial one has $\beta=0$.},
\begin{eqnarray}
g_L(x) \in U_L, &\;& g_R(x) \in U_R \nonumber \\
q_L &\rightarrow& g_L q_L \nonumber \\
q_R &\rightarrow& g_R q_R \nonumber \\
l_\mu &\rightarrow& g_L l_\mu {g}_L^{\dagger} + 
i g_L \partial_\mu {g}_L^{\dagger} \nonumber \\
r_\mu &\rightarrow& g_R r_\mu {g}_R^{\dagger} + 
i g_R \partial_\mu {g}_R^{\dagger} \nonumber \\
s+ip &\rightarrow& g_R \; (s+ip) \; {g}_L^{\dagger} \; ,
\label{2.8}
\end{eqnarray}
na\" {\i}vely becomes a local gauge symmetry for 
the lagrangian density in eq. (1). 
This is not quite so, due to the fact that the transformations in (\ref{2.8})
also induce a non-trivial
anomalous  $U_A(1)$
transformation on the generating functional $Z[l, r, s, p, \theta]$. Its
origin may be traced to the transformation properties of the fermionic 
integration measure \cite{f} (or, if one wishes, of the fermion determinant)
once it is properly regularized. This is  reflected in the
 anomalous divergence in the ninth axial
singlet current,
\begin{equation}
\label{2.9}  
\partial_\mu J_5^{\mu \; (0)}=  \frac {g^2}{16 \pi^2} \frac {1}{N_c}
Tr_c \left({G_{\mu \nu} {\tilde{G}}^{\mu \nu}} \right)
\;;\;\;\;\; J_5^{\mu \; (0)}=\bar{q} \gamma_\mu\gamma_5  q.
\end{equation}
  This anomaly-related effect turns off any  
potential advantage inherent to the existence of a gauge symmetry which
eventually would severely constraint the form of the effective action. 
However, this drawback is obviated as the   
{ $U_A(1)$ anomaly contribution may be
altogether eliminated by 
 judiciously choosing the transformation law for the external 
field $\theta(x)$. Indeed, 
for infinitesimal $g_L=I + i (\beta - \alpha)$,
$g_R=I+ i (\beta + \alpha)$,  the source $\theta (x)$ ought to change as
$$\theta(x) \rightarrow \theta (x) - 2 \langle \alpha(x)\rangle ,$$
(here we have switched to the standard notation and denote the trace
operation over flavour indices by brackets $tr_F (...) \equiv \langle  ... \rangle $)
in order for the $U_A(1)$ anomaly to cancel. The term generated by the 
anomaly in the fermion determinant is explicitly compensated by
the shift in the $\theta$ source.  

A subtlety is still to be analized. The set of local gauge 
transformations we have
constructed in (\ref{2.8}) also induces a non-abelian anomaly. This
new drawback can not be circumvented and needs explicit consideration.
As discussed in \cite{b}, imposing upon regularization the
requirement of conservation for the nine
vector currents, the change in $Z$ under (\ref{2.8}) reads 
$$
\delta Z \equiv - \int dx \; \langle  \alpha(x) \Omega(x) \rangle , $$ 
$$
\Omega(x)=\frac {N_c}{16 \pi^2} \epsilon^{\alpha \beta \mu \nu}
\left[ v_{\alpha \beta} v_{\mu \nu} + \frac {4}{3} (\nabla_{\alpha}
a_{\beta}) (\nabla_{\mu} a_\nu) + \frac {2}{3} i \{ v_{\alpha \beta},
a_\mu a_\nu \} + \frac {8}{3} i a_\mu v_{\alpha \beta} a_\nu +
+ \frac {4}{3} a_\alpha a_\beta a_\mu a_\nu \right] \;, $$
\begin{equation}
\label{2.11}
v_{\alpha \beta}= \partial_\alpha v_\beta - \partial_\beta v_\alpha
- i [v_\alpha, v_\beta] \;, \;\;\;\;\;\;\;\;\;\
\nabla_\alpha a_\beta=\partial_\alpha a_\beta - \partial_\beta a_\alpha
- i [v_\alpha, a_\beta].
\end {equation}
The terms in $\delta Z$ appear due to triangle AVV Feynman diagrams which
involve insertions of an axial and two vector quark bilinear operators.
Higher polygon-shaped diagrams, quadrangles and pentagons are anomalous
as well and give rise to the cubic and quartic terms in (\ref{2.11}). 
Adler and Bardeen showed that the coefficients of the
anomaly are not affected by higher-order radiative corrections, i.e.,
diagrams with more than one-loop do not contribute to the anomalous terms
\cite{ab}. $\delta Z$  fulfills the Wess-Zumino consistency conditions 
\cite{wz}.

The integrated form of this anomaly must be added by hand to the
low-energy effective field theory that we shall construct. Such a theory
will consist of interacting bosons and therefore will not be anomalous. 
Thus, the above effect will be contained in an additional term to
the effective lagrangian.

It is worth pointing out that unlike for $SU_L \otimes SU_R$,
here there is the possibility of combining one Wess-Zumino-Witten (WZW) vertex
(which start out at $O(p^4)$) with the $O(p^0)$ pieces in the lagrangian
and generate additional one-loop divergences at $O(p^4)$. They will
be given elsewhere \cite{anomalia}. Notice that 
for $SU_L \otimes SU_R$ this cannot occur at $O(p^4)$ because
the lowest chiral power counting are $O(p^2)$ terms, which generate 
divergences at $O(p^6)$.

%

\section{The chiral lagrangian} 
The mere knowledge of the symmetries and the way they are realized
provide an enormous insight for they are reflected in every
aspect of the theory, e.g., in
the spectrum, the Green's functions, the interactions, to mention a few.
In the case of the strong interactions the symmetries are the bulk
of the information which is available at low energies,
since QCD is non-perturbative there. Although a lot of effort has been put
in numerical calculation projects and enormous progress in solving
the technical difficulties has been achieved, so far 
the problem has remained too hard to tackle satisfactorily. 

In this section we shall write an effective lagrangian for the soft
interactions of the
lightest particles in the spectrum, the nonet of pseudoscalar mesons, 
$\pi$, $K$, $\eta$, $\eta'$.

The effective lagrangian is a combined statement about the degrees
of freedom and the symmetries that are relevant
for the processes under study.
{\it Effective} refers to the choice of field variables,
in this case fields for the $\pi$, $K$, $\eta$, $\eta'$ particles that are 
observed in the range of energies below the $\rho$-meson mass $m_{\rho}$.
More generally, being the lightest particles in the spectrum 
the use of the effective lagrangian will be the determination of
the long distance behaviour of any of 
the QCD Green's functions, where they are expected to dominate.

Being a symmetry statement, the strategy consists of writing down 
for the effective lagrangian the most general expression  
that contains all the independent terms compatible with the symmetries, 
multiplied by unknown constants.
The fate of many effective theories
is to be of little practical use if the number of unknown constants
blows up. They can always be fitted from experiment but too often the
number of experimental results available is of the same order 
(if not smaller) as that
of constants, rendering the approach little predictive.

In the present case the spontaneously broken symmetry character 
and the consequent goldstone nature of the $\pi$, $K$, $\eta$, $\eta'$
impose severe constraints on the form of the interactions. The 
number of unknown constants reduces in a drastic manner
if one restricts to the lower terms. Actually, they would reduce to a handful
of them if there were no $U_A(1)$ anomaly effects to incorporate, as happens
in the $SU_L \otimes SU_R$ lagrangian if one only keeps terms up
to $O(p^4)$. Fortunately, the
swarm of new terms that the $U_A(1)$ anomaly introduces
carry high powers of $1/N_c$, and to lower orders in $1/N_c$ 
only a few survive.

Following \cite{gl}, \cite{l} we collect the nine pseudoscalar fields
in a hermitian matrix $\Phi (x)$, 
$$\Phi (x)= \eta^0(x) \lambda_0+ \pi (x)$$
where $\pi(x)=\vec{\pi}(x) \cdot \vec{\lambda}$, 
$\lambda_0= \sqrt{\frac {2}{n_l}} I$ and 
$\vec{\lambda}$ are the Gell-Mann matrices of $SU(n_l)$. For $n_l=3$ (see
Appendix B),
\begin{equation}
\label{3.1}
\Phi (x)=\left(\begin{array}{ccc} 
\frac{1}{\sqrt{2}}\pi^0 + \frac{1}{\sqrt{6}}\eta^8+ \sqrt{\frac {2}{3}} \eta^0
 & \pi^+ & K^+ \\
\pi^- & -\frac{1}{\sqrt{2}}\pi^0+ \frac{1}{\sqrt{6}}\eta^8+ 
\sqrt{\frac {2}{3}} \eta^0 & K^0 \\
K^- & {\bar{K}}^0 & - \frac{2} {\sqrt{6}}\eta^8+ \sqrt{\frac {2}{3}} \eta^0
\end{array}\right).
\end{equation}
The unitary $n_l \times n_l$ matrix  $U(x)$ is the exponential of $\Phi (x)$,
\begin{equation}
U(x) \equiv e^{ i \Phi (x) / f }\ ,
\label{U}
\end{equation}
$f$ is an order parameter of chiral symmetry breaking and gives the strength
of the coupling between 
the goldstone bosons and the currents that are spontaneously broken
and do not annihilate the vacuum.
In this case $\det U$ is a phase, 
$$\det U= \exp ( i \sqrt{2 n_l} \; \eta^0 /f ).$$

Under $U_L \otimes U_R$, U transforms linearly,
\begin{equation}
\label{3.2}
U \rightarrow g_R U g^{\dagger}_L.
\end{equation}

The transformations induced by (\ref{3.2}) on $\Phi$ are more involved.
Under $U_{R+L}(1)$, $\Phi$ is automatically invariant, as it should, 
for mesons must carry baryon number equal to zero.
Under $SU_{L+R}(3)$, $\Phi$ transforms linearly: it contains
two irreducible representations, the octet $\pi$ and the singlet
$\eta^0$. 
Under an axial transformation
$$U \longrightarrow e^{i \alpha} U e^{i \alpha},$$
$\Phi$ changes nonlinearly,
$$ \Phi \rightarrow \Phi + 2 \alpha + O(\alpha^2).$$
This can be understood on geometrical grounds
since the fields in $\Phi$ may be regarded as coordinates that span
the coset space $U_L \otimes U_R \; / \; U_{R+L}$, upon which
the $U_L \otimes U_R$ acts: the fields
themselves are, in a sense, parameters of a group element,
and the nonlinearity 
reflects the group transformation law when written in terms of the
continuous parameters. 
For $\langle \log U \rangle = i \sqrt{2 n_l} \eta^0/f$,
$$\langle  \log U\rangle  \
\longrightarrow \ \langle  \log 
U\rangle  + \langle  \log (g_R g_L^{\dagger})\rangle = \langle  \log U\rangle 
+ 2 i \langle \alpha\rangle \; ;$$
$\eta^0$ gets thus shifted only under an axial transformation and is invariant
under any vector one.
Under $U_L \otimes U_R$ it never mixes with any of the $\pi$ components.

With this choice of fields, the origin of the derivative couplings among
goldstone-boson becomes transparent: a lagrangian
invariant under global $U_L \otimes U_R$ ought to
reduce to zero if $U$ were a constant matrix, for then, by
virtue of the symmetry, it could always be transformed away with a global
unitary rotation: the couplings need thus
be derivative couplings \cite{llibre de Georgi}. 
The expansion will be in powers of momenta of the soft light mesons divided
by a scale of chiral symmetry breaking, which is of order $ \sim 4 \pi f \;
\sim m_\rho$, the mass of the next excited state, the $\rho$-meson. 
In addition, in the present case the $U_A (1)$ anomaly introduces 
novel couplings of the $\eta_0$ meson that are not derivative couplings.

The objects at hand to construct the chiral
effective lagrangian are the matrix $U$ and the external sources,
$r_\mu$, $l_\mu$, $(s+ip)$ and $\theta$. 
It is useful to introduce covariant derivatives for the fields 
and the sources. 
The covariant derivative will act on flavour space and, as usual,
its action will depend on the transformation law of the object it derives,
\begin{eqnarray}
D_\mu U &=& \partial_\mu U  - i r_\mu U + i U l_\mu, \nonumber \\
D_\mu (s+ip) &=& \partial_\mu (s+ip)  - i r_\mu (s+ip) + 
i  (s+ip) l_\mu, \nonumber \\
D_\mu \langle \log  U\rangle &=&\langle U^\dagger D_\mu U\rangle  = 
\partial_\mu \langle \log  U\rangle  - 
i \langle  r_\mu - l_\mu\rangle , \nonumber \\
i D_\mu \theta &=& i \partial_\mu \theta + i \langle r_\mu - l_\mu\rangle .
\label{3.3}
\end{eqnarray}
One is led to introduce field strengths for the vector
and axial sources
\begin{eqnarray}
F^L_{\mu \nu}= \partial_\mu l_\nu - \partial_\nu l_\mu - i [ l_\mu, l_\nu],
\nonumber \\
F^R_{\mu \nu}= \partial_\mu r_\nu - \partial_\nu r_\mu - i [ r_\mu, r_\nu].
\label{3.4}
\end{eqnarray}
Under local $U_L \otimes U_R$ they transform as
\begin{eqnarray}
D_\mu U &\rightarrow& g_R 
\left( D_\mu U  \right) g_L^{\dagger} \nonumber \\
D_\mu (s+ip) &\rightarrow& g_R \left(
D_\mu (s+ip)  \right) g_L^{\dagger} \nonumber \\
F^L_{\mu \nu} &\rightarrow& g_L F^L_{\mu \nu} g_L^{\dagger} \nonumber \\
F^R_{\mu \nu} &\rightarrow& g_R F^R_{\mu \nu} g_R^{\dagger}.
\label{3.5}
\end{eqnarray}

The combination 
\begin{equation}
X(x) \equiv \langle \log  U(x)\rangle +\hat{\theta}(x) = 
i\ \frac{\sqrt{2\nz}}{f}\ \eta^0 +\hat{\theta}(x) , \;\;\;\;
\hat{\theta} \equiv i \theta,
\label{3.5bis}
\end{equation}
is invariant, and so is any function of $X$ \cite{w1}, \cite{l}.
This is a novelty of
$U_L \otimes U_R$ and it is possible because 
$\langle \log U \rangle$ does not vanish, as it does for $SU_L \otimes
SU_R$. Due to the $U_A(1)$ anomaly, each invariant operator 
generates, in reality, an infinite set of invariant operators, since
the symmetry allows to multiply it by any function of $X$ and still remain
invariant. Notice that this method of finding new operators by multiplying
the old ones by functions of $X$ never introduces new derivatives
to the vertices. It is for the same reason that counting the number
of derivatives and retaining operators that contain up to a certain 
number does not limit the number free constants, as it used to
in $SU_L \otimes SU_R$. At each order in the derivative expansion
we find an infinite number of constants.

It is customary to introduce the source $\chi (x)$
$$\chi \equiv 2 B (s+ip),$$
which transforms as the $U$ matrix.
$B$ is a constant that is related to the quark condensate 
$\langle \bar{q} q\rangle$.
Its relevance comes from the fact that the explicit symmetry breaking driven
by the quark masses provides the former goldstone bosons also with a mass.
If $m_q$ denotes a quark mass, 
to lowest chiral order there is a contribution to the meson masses
which involves $B$ and is given by $m_\pi^2= 2 B m_q$. We {\it assume} that
it is the bulk of it: we make the {\it hypothesis} that
further pieces which involve order parameters other
than $B$ are negligible, comparatively.
This leads to counting $\chi$ as $O(p^2)$.
The $\eta'$ gets an additional piece to its mass through the $U_A(1)$ anomaly,
which is $O(1/N_c)$, and which does not add to the masses in the octet
(\ref{mass}).
The singlet-octet mass splitting is very big in practice, of $ \sim 400$ MeV
in the least extreme case.

All the constants that appear in the effective theory are free, to
be fitted from experiment. Although we have not been able to compute
them from QCD, there are exact inequalities that have to be verified, which are
based on the vector structure of QCD. These relations are non-perturbative
\cite{nosaltres}.

\vspace*{0.5cm}

From the effective lagrangian, one can make contact with QCD through
the generating functional $Z[l,r,s,p,\theta]$ introduced in section 2.
One demands that the same functional - the same Green's functions -
should be obtained by starting from the effective theory as from QCD.
It can be formally
written in terms of the light pseudoscalar fields, collected in $U(x)$, as
\begin{equation}
\label{3.6} \left.
exp \{ i Z[l,r,s,p,\theta] \}= \int [dU] \; e^{ i \int dx \; L_{eff} }
\;\;\;   \right|_{low \; modes}.
\end{equation} 
The chiral lagrangian only copes with the
long distance behaviour of Green's functions. Only the lowest modes have 
physical significance in (\ref{3.6}). For distances smaller than 
$1/m_\rho$ the approach is inappropriate and
the integrals over loop momenta have a natural cutoff associated with them
that is $m_\rho$. The higher modes, corresponding to more energetic
pseudoscalars, and the rest of massive hadrons are integrated out and
their effect manifests through the coupling constants of the effective
theory \cite{e}.

The most general lagrangian invariant under (local) $U_L \otimes U_R$
\footnote{Terms of the sort $\sum_{\alpha=0}^{8}\langle \lambda^\alpha
U^\dagger D_\mu U\rangle \langle \lambda^\alpha U^\dagger D^\mu U\rangle $ 
and alike, given the 
properties of the $\{ \lambda^\alpha \}$ matrices, can be re-expressed
in terms of the operators written in the text. In this particular case, with
the relation $\sum_{\alpha=0}^{8} \lambda^\alpha_{ij} \lambda^\alpha_{kl}=
2 \delta_{il} \delta_{jk}$, it becomes $-2\langle  
(D_\mu U^\dagger)( D^\mu U)\rangle $}
that includes terms with two derivatives or less, or one
power of $\chi$, is the following \cite{gl}, \cite{l}, 
\begin{eqnarray}
{\cal L}_{(0+2)}=&-&W_0(X)+ W_1(X) \langle D_\mu U^\dagger D^\mu U\rangle  +
W_2(X) \langle U^\dagger \chi + \chi^\dagger U\rangle  + i W_3(X)
\langle U^\dagger \chi - \chi^\dagger U\rangle  \nonumber \\
&+& W_4(X) \langle U^\dagger D_\mu U\rangle \langle U^\dagger D^\mu U\rangle +
 W_5(X) \langle U^\dagger (D_\mu U)\rangle  D^\mu \hat{\theta} 
+ W_6 (X)  D_\mu \hat{\theta} D^\mu \hat{\theta}.
\label{3.7}
\end{eqnarray}

Parity conservation implies that they 
are all even functions of $X$ except for $W_3$ that is odd. Furthermore,
$W_4(0)=0$, $W_1(0)=W_2(0)=\frac {f^2}{4}$ gives the correct normalization 
for the quadratic terms.

The first term $W_0$ has neither derivatives nor powers of $\chi$, and
therefore counts as $O(p^0)$. The rest of the terms count as $O(p^2)$.
The $1/N_c$ power counting of the diverse couplings is given in the next
section.
The term $W_0$ brings a mass terms for the $\eta^0$
\begin{equation}
\label{mass} 
\left. m_{\eta^0}^2 \right|_{U_A(1)} = - \frac {2 n_l}{f^2} W_0''(0),
\end{equation}
whereas the first term in the expansion of $W_3$ gives a contribution
to singlet-octet mixing from $U_A(1)$.

The $N_c \to \infty$ limit of QCD actually imposes more restrictions
on the effective lagrangian than the symmetry $U_L \otimes U_R$ alone.
Under  $U_L \otimes U_R$, the fields
$\eta^0$ and $\pi$ never mix their components.
There is no reason why the particles created by them should bear
any sort of relationship whatsoever, as though 
they belonged to the same irreducible 
representation, like the $\pi$ do. There is no reason, either, why in
the definition  of $U(x)$ in (\ref{3.1}) $\eta^0$ and $\pi$ 
should appear in the exponent
divided by the same constant $f$, i.e., normalised 
in the same way. One could have written instead
$$U(x)= e^{ i \left( \sqrt{\frac {2}{n_l}} \frac {\eta^0}{f_0} +
\frac {\pi}{f} \right)}\ ,$$
with $f$ and $f_0$ unrelated and, yet, (\ref{3.7}) would be 
$U_L \otimes U_R$ invariant. 
The proper way to cast this issue requires to fix the normalization
of each field by looking at the kinetic energy too, not at $U(x)$ only. 
In (\ref{3.7}), the kinetic energy terms are
$ \frac {f^2}{4} \langle D_\mu U^\dagger D^\mu U\rangle 
+W_4(0) \langle U^\dagger D_\mu U\rangle \langle U^\dagger D^\mu U\rangle ,$ 
which read,
$$ \frac {1}{2} \left(\partial_\mu \vec{\pi} \right)
\cdot \left( \partial^\mu \vec{\pi} \right)+\frac {1}{2} \left(
\frac {f^2}{f_0^2}
-\frac{ 4 n_l W_4(0)}{f_0^2} \right) \partial_\mu \eta_0 
\partial^\mu \eta_0.$$
The normalization condition $\left(f^2-4 n_l W_4(0) \right)/ f_0^2 =1$ 
relates three constants;
it may be viewed as an arbitrariness in their definition, for a change in 
$f_0$ can be always compensated for by an appropriate change in $W_4(0)$. 
Once this normalization is fixed,
the strength that the singlet $\eta^0$ couples to the singlet axial
current is $f_0$ whereas the octet couple
to the octet axial current with strength $f$. 

Unlike $U_L \otimes U_R$, which does not have a dimension nine irreducible
representation, the $N_c \to \infty$ 
really enforces a nonet symmetry, with the $\pi$, $K$, $\eta$, $\eta '$
all having identical properties. 
Indeed, the planar Feynman diagrams that contribute
to $\langle J_{5 \ \mu}^{(0)}(x) J_{5 \ \nu}^{(0)}(0)\rangle $ and to 
$\langle J_{5 \ \mu}^{(a)}(x) J_{5 \ \nu}^{(a)}(0)\rangle $ ($a=1,...,8$), with
$J_{5 \ \mu}^{\rho}=\bar{q} \gamma_\mu \gamma_5 \frac {\lambda^\rho}{2} q$,
are the same since the $\bar{q} q \to gluon$ annihilation diagrams that
would only contribute to the singlet channel turn out to be
$1/N_c$ suppressed (OZI violating processes).
Barring for $1/N_c$ corrections the two decay constants coincide $f/f_0=1$.
Moreover, if the same quark mass were switched on for all light quark species, 
a mass would be generated identical for the singlet $\eta_0$ as for the 
octet particles \cite{w1} \cite{v}. 

The standard normalization of the kinetic energy and
nonet symmetry require $W_4(0)=0$.
Without loss of generality, in that case
the term $W_4(X)$ may be eliminated altogether
by a change of field variables of the type
$U \to U \; \exp [ i F(X) ]$, that maintains the transformation properties
for $U$ (but changes the normalization conditions of $\eta^0$).

\vspace*{0.5cm}
The methods of the background field and the steepest descent, when applied
to the functional integral (\ref{3.6}), provide
the loop-wise expansion of the generating functional.
The Green's functions are read off from
the generating functional, about its minimum when the external
sources are switched off, $\chi = 2 B \; diag (m_u, \; m_d, \; m_s)$ and 
$\hat{\theta}=0$.
To lowest order we assume that the minimum is achieved at $U_0=I$, which is
compatible with the equation that minimizes the lowest order effective
action \cite{gl}.

In order to include one-loop corrections, 
one proceeds to introduce a background
field  matrix $\bar{U} (x)$, and expand (\ref{3.7})
about this background configuration. For that we decompose $U(x)$ as
$$U(x)=\bar{U}(x) \Sigma (x), \;\;\;\;\;\;\;\;\; \Sigma(x) =e^{i\Delta(x)}\ ,$$
with $\Delta$ the matrix of quantum fields
\footnote{The procedure is not manifestly {\it left - right} symmetric. This
is not a worrisome issue given that the quantization of scalar fields in four
dimensions does not have any anomaly that would favor one over the other.
Simplicity reasons have lead to our choice. A more symmetric treatment
would lead to the same final results.}.
By choosing $\bar{U}$ to transform under $U_L \otimes U_R$ as $U$,
the transformation laws for $\Sigma$ and $\Delta$ become

$$\Delta \rightarrow g_L  \Delta  g_L^\dagger , \;\;\;\;\;\;
\Sigma  \rightarrow g_L  \Sigma  g_L^\dagger,$$
and, therefore, their covariant derivatives are
$$D_\mu \Sigma = \partial_\mu \Sigma -i[l_\mu,\Sigma ], \;\;\;\;\;\;
D_\mu \Delta = \partial_\mu \Delta -i[l_\mu,\Delta ].$$
 
Next we expand the lagrangian in (\ref{3.7}) about $\bar{U}$ and
keep terms up to quadratic in $\Delta$. The one-loop effective action
is obtained upon evaluation of the functional determinant of the
differential operator that appears in the piece quadratic in $\Delta$. 
Finally the background
field $\bar{U}(x)$, which is {\it a priori} independent of the sources,
is judiciously chosen so as to verify the equations
$\delta Z[\bar{U},l,r,s,p,\theta] =0$ with the sources held fixed.
The method ensures that with this $\bar{U}$, Z is the effective action
$\Gamma[\bar{U}]$ -  the generator of the one-particle irreducible Green's 
functions of fields gathered in $U$ -, in the presence of the external 
sources.
For the one-loop effective action it suffices to use
the tree level equations of motion.
The corrections to it would modify the two-loop effective action.

Expanding the effective lagrangian in (\ref{3.7}) about $\bar{U}$,
disregarding the terms linear in $\Delta(x)$ that vanish with the
equations of motion, it yields:
\begin{eqnarray}
{\cal L}_{0+2}(\bar{U} \Sigma)&=&{\cal L}_{0+2}(\bar{U})+
{\cal A} \; \langle \Delta\rangle ^2 \nonumber \\
&+& 2 W_1'(\bar{X}) \langle \Delta\rangle  \langle C^\mu 
\ D_\mu \Delta\rangle  
+ W_1(\bar{X}) \langle D_\mu \Delta  \ D^\mu \Delta\rangle  
+ W_1(\bar{X}) \langle C_\mu [\Delta, D_\mu \Delta]\rangle  \nonumber \\ 
&+& W_2'(\bar{X}) \langle \Delta\rangle  \langle \Delta N\rangle  
- \frac {1}{2} W_2 (\bar{X}) \langle \Delta ^2 M\rangle  \nonumber \\
&+& i W_3'(\bar{X}) \langle \Delta\rangle  \langle \Delta M\rangle  - 
\frac {i}{2} W_3 (\bar{X}) 
\langle \Delta ^2 N\rangle  \nonumber \\
&-& W_5' (\bar{X}) \langle \Delta\rangle \langle D_\mu \Delta\rangle  
D^\mu \hat{\theta}
+O(\Delta)^3 ,
\label{3.8}
\end{eqnarray}
where
\begin{eqnarray}
{\cal A} &\equiv& \frac {1}{2} W_0''(\bar{X}) + \frac {1}{2} W_1''(\bar{X}) 
\langle C_\mu C^\mu\rangle  - \frac {1}{2} W_2''(\bar{X}) \langle M\rangle  
\nonumber \\
&-& \frac {i}{2} W_3''(\bar{X}) \langle N\rangle  - \frac {1}{2} W_5''(\bar{X})
\langle C_\mu\rangle  D^\mu \hat{\theta} - \frac {1}{2} W_6''(\bar{X})
D_\mu \hat{\theta} D^\mu \hat{\theta},
\label{3.9}
\end{eqnarray}
and
$$C_\mu \equiv \bar{U}^\dagger D_\mu \bar{U},\;\;\;\;\;\ 
M \equiv \bar{U}^\dagger \chi + \chi^\dagger \bar{U},\;\;\;\;\;
N\equiv \bar{U}^\dagger \chi - \chi^\dagger \bar{U};$$
all transform as $C_\mu$, $M \to g_L M g_L^\dagger$,
$N \to g_L N g_L^\dagger$; also, $C_\mu^\dagger= - C_\mu$, $M^\dagger = M$
and $N^\dagger=-N$.
The functions $W_i (X)$ have been Taylor expanded about the background value
$\bar{X}=\langle \log  \bar{U}\rangle  + \hat{\theta}$ as follows,
$$W_k (\langle  \log (\bar{U} \Sigma)\rangle + \hat{\theta})=
W_k (\bar{X}) + i W_k'(\bar{X}) \langle \Delta\rangle 
-\frac {1}{2} W_k''(\bar{X}) \langle \Delta\rangle ^2 + O(\Delta)^3.$$
Integrations by parts have been performed where necessary and the
total divergences have been discarded.
The equations of motion can be read off from the terms that are 
linear in the variation $\Delta$,
\begin{eqnarray}
D_\mu C^\mu &=& \frac {1}{2} \frac {W_0'}{W_1} +
\frac {1}{2} \frac {W_1'}{W_1} \langle C^\mu C_\mu\rangle 
- \frac {W_1'}{W_1} (\langle C_\mu\rangle +D_\mu \hat{\theta}) C^\mu 
\nonumber \\
&+&\frac {1}{2} \frac {W_2}{W_1} N + \frac {i}{2} \frac{W_3}{W_1} M
- \frac {1}{2} \frac {W_2'}{W_1} \langle M\rangle  - \frac{i}{2} 
\frac {W_3'}{W_1} \langle N\rangle 
\nonumber \\
&+& \left( \frac {1}{2} \frac {W_5'}{W_1}- \frac {1}{2} 
\frac {W_6'}{W_1}  \right)
(D_\mu \hat{\theta})(D^\mu \hat{\theta})
+ \frac {1}{2} \frac {W_5}{W_1} D_\mu D^\mu \hat{\theta},
\label{3.10}
\end{eqnarray}

(Henceforth the arguments $\bar{X}$ are omitted from the functions
$W_k$'s and their derivatives that appear in the calculation; 
also, bars are suppressed from $\bar{X}$ and $\bar{U}$).

In order to be able to use the expressions collected in the Appendix A for the
evaluation of the one-loop divengent parts
it proves useful to perform a change of
integration variables
so as to leave the operator that acts on the quadratic piece in
(\ref{3.8})
in the usual form, with the laplacian piece $\partial_\mu \partial^\mu$
multiplied by a constant, not by a function. 
For this purpose we change variables to
\begin{equation}
\Delta(x)=  \frac {f}{2} \frac {1}{\sqrt{W_1}} \varphi (x),
\label{3.11}
\end{equation}
and expand the hermitian matrix $\varphi(x)$ in the basis of matrices
$\lambda^\alpha$ (see Appendix B):
$$\varphi = \varphi^\alpha \lambda^\alpha, \;\;\;\;\;\;\;\;\varphi^\alpha=
\frac {1}{2}\langle \varphi \lambda^\alpha\rangle .$$
Retaining the piece quadratic in the quantum fluctuating fields 
$\varphi^\alpha$, one finds
\begin{equation}
{\cal L}_{(0+2)}^{Quadratic}= -\frac{f^2}{2} \varphi^\alpha 
\left( d_\mu d^\mu + \sigma \right)^{\alpha \beta} 
\varphi^\beta ,
\label{3.12}
\end{equation}
where
\begin{equation}
[d_\mu \varphi]^\alpha= \partial_\mu \varphi^\alpha + \omega_\mu^{\alpha
\beta} \varphi^\beta,
\label{3.13}
\end{equation}
and 
\begin{equation}
\omega_\mu^{\alpha \beta}=
\frac {i}{2} \langle (l_\mu + \frac {i}{2} C_\mu)
[\lambda^\alpha,\lambda^\beta]\rangle 
+\frac {1}{4} \frac {W_1'}{W_1} \left(\langle C_\mu 
\lambda^\alpha\rangle \langle \lambda^\beta\rangle 
-\langle C_\mu \lambda^\beta\rangle \langle \lambda^\alpha\rangle  \right).
\label{3.14}
\end{equation}
Notice that $\omega_\mu^{\alpha \beta}$ is antisymmetric in
$\alpha$, $\beta$. It is this property what allows to integrate $d_\mu$
by parts as a whole, as though it were the single $\partial_\mu$.
For that reason the operator $d_\mu d^\mu + \sigma$ is manifestly
hermitian. The expression in (\ref{3.12}) differs from that of
(\ref{3.8}), after the change of variables, by a total derivative. The 
evaluation of the Gaussian integral involves the expression 
(\ref{3.12}), though: 
it is the determinant of the differential operator that is
{\it hermitian} the one that has to be evaluated
\footnote{In practice this means that any term of the sort $\varphi^\alpha
f^{\alpha \beta}_\mu (x) \partial^\mu \varphi^\beta$ (which,
in general, is {\it not} hermitian as can be seen 
if one tries to bring the operator act on
$\varphi^\alpha$, on the left) can always be written as $\varphi^\alpha
a^{\alpha \beta}_\mu (x) \partial^\mu \varphi^\beta- \frac {1}{2} 
\varphi^\alpha
\partial_\mu \left( s^{\alpha \beta}_\mu (x) \right) \varphi^\beta
+\frac {1}{2} \partial_\mu \left( 
\varphi^\alpha s^{\alpha \beta}_\mu (x) \varphi^\beta  \right)$. Notice
that the last term is a total derivative which we shall discard. The
remaining part
$a^{\alpha \beta}_\mu(x) \partial^\mu$
is hermitian now. The $a^{\alpha \beta}_\mu (x)$ 
and $s^{\alpha \beta}_\mu (x)$ are the antisymmetric and symmetric parts
of $f^{\alpha \beta}_\mu (x)$, respectively. This decomposition
is unique.}.

The curvature associated to this connection is
\begin{eqnarray}
R^{(\alpha \beta)}_{\mu \nu}&=& \partial_\mu \omega_\nu^{\alpha \beta} -
\partial_\nu \omega_\mu^{\alpha \beta}+
\omega_\mu^{\alpha \gamma} \omega_\nu^{\gamma \beta} -
\omega_\nu^{\alpha \gamma} \omega_\mu^{\gamma \beta} \nonumber \\
&=&\frac {i}{4} \langle  Q_{\mu \nu} [\lambda^\alpha, \lambda^\beta] \rangle +
\frac {1}{4} \left( \langle H_{\mu \nu} \lambda^\alpha\rangle \langle 
\lambda^\beta\rangle 
-\langle H_{\mu \nu} \lambda^\beta\rangle \langle \lambda^\alpha\rangle  
\right) \nonumber \\
&-& \frac{n_l}{8} \left( \frac {W_1'}{W_1} \right)^2
\left( \langle C_\mu \lambda^\alpha\rangle \langle C_\nu \lambda^\beta\rangle 
-\langle C_\mu \lambda^\beta\rangle \langle C_\nu \lambda^\alpha\rangle 
\right),
\label{3.14p}
\end{eqnarray}
$\mu, \nu$ are space-time indices, whereas $\alpha, \beta, \gamma$
label the Gell-Mann matrices of $U(n_l)$,
$$Q^{\mu \nu}= F_L^{\mu \nu} + U^\dagger F_R^{\mu \nu} U - \frac {i}{2}
[C^\mu, C^\nu],$$
and
\begin{eqnarray}
H^{\mu \nu} &=& \left( \frac {W_1''}{W_1} - \frac {3}{2} 
\left(\frac {W_1'}{W_1}\right)^2 \right) 
\left( \langle C^\mu\rangle  C^\nu - \langle C^\nu\rangle  C^\mu \right)
+\left( \frac {W_1''}{W_1} - \left(\frac {W_1'}{W_1} \right)^2 \right)
\left( C^\nu D^\mu \hat{\theta} - C^\mu D^\nu \hat{\theta} \right)
\nonumber \\
&+& \frac {W_1'}{W_1} \left( i F_L^{\mu \nu} - i U^\dagger F_R^{\mu \nu} U
\right). \nonumber
\end{eqnarray}
For $\sigma$ we find
\begin{eqnarray}
\sigma^{\alpha \beta} &=& \frac{1}{8} \langle [C_\mu, \lambda^\alpha]
[C^\mu, \lambda^\beta]\rangle +\frac{1}{8} \langle 
R \{\lambda^\alpha,\lambda^\beta \}\rangle 
+S \delta^{\alpha \beta} 
+\frac {n_l}{8} \left( \frac {W_1'}{W_1} \right)^2 \langle C_\mu 
\lambda^\alpha\rangle 
\langle C^\mu \lambda^\beta\rangle  \nonumber \\
&+& \frac {1}{4} ( \langle T \lambda^\alpha\rangle \langle 
\lambda^\beta\rangle 
+\langle T \lambda^\beta\rangle \langle \lambda^\alpha\rangle  ),
\label{3.15}
\end{eqnarray}
where
\begin{eqnarray}
S &=& -\frac {1}{2} \left( \frac {W_1''}{W_1} - \frac {1}{2} \left(
\frac{W_1'}{W_1} \right)^2 \right) (\langle C_\mu\rangle + 
D_\mu\hat {\theta} \;)^2
-\frac {1}{2} \frac {W_1'}{W_1} (\langle D_\mu C^\mu\rangle  + 
D_\mu D^\mu \hat{\theta}\;), \nonumber \\
T &=& -\frac {1}{2} \frac {W_0''}{W_1} +
\left( \frac {W_1''}{W_1} - \frac {1}{2} \left(
\frac{W_1'}{W_1} \right)^2 \right) ( \langle C_\mu\rangle  C^\mu - \frac {1}{2}
\langle C_\mu C^\mu\rangle  ) + \frac {W_1'}{W_1} (D_\mu C^\mu) \nonumber \\
&+& \frac {1}{2} \langle  \frac {W_2''}{W_1}M + i \frac {W_3''}{W_1} N 
\rangle  - 
\frac {W_2'}{W_1} N - i \frac{W_3'}{W_1} M +
\frac {W_1''}{W_1} C^{\mu} (D_\mu \hat{\theta}) - 
\frac {1}{2}\frac {W_5'}{W_1} D_\mu D^\mu \hat{\theta}
\nonumber \\ &+&\frac {1}{2} \left( \frac {W_6''}{W_1} -
 \frac {W_5''}{W_1} \right)(D_\mu \hat{\theta})^2,
\nonumber \\ 
R &=& \frac {W_2}{W_1} M+i \frac {W_3}{W_1} N.
\label{3.16}
\end{eqnarray} 

The one-loop effective action is obtained by including the quadratic 
fluctuations
about the configuration $[\bar{U}]$ that, consistently, minimizes the
effective action itself. One needs to evaluate the integral
of a Gaussian functional, with the known formal result
$$\int [d \varphi] e^{- i \frac{f^2}{2}  \int d^4x \; \varphi^\alpha
( d_\mu d^\mu + \sigma)^{\alpha \beta} \varphi^\beta }
\sim \frac {1}{\sqrt{\det \left( d_\mu d^\mu+ \sigma \right)}},$$
which, upon exponentiation, contributes to the effective action as
\begin{equation}
\Gamma^{One-loop}_{eff}[\bar{U}]=\int d^4x \;  {\cal L}_{(0+2)}(\bar{U})  
+\frac{i}{2} Tr \; \log \left(d_\mu d^\mu + \sigma \right)+
\int d^4x \; {\cal L}_{(4)}(\bar{U}).
\label{3.17}
\end{equation}

A word is needed on the proper definition of the previous expressions.
The heat-kernel technique has been used to define the determinant
(see Appendix A), and the divergences have been dealt with dimensional
regularization. 
In order to absorb the infinities that result from the functional determinant,
counter-terms of $O(p^2)$ and $O(p^0)$, as well as new terms of $O(p^4)$ 
need be included. 

The determinant from the change of functional integration variables in
(\ref{3.11}) gives a contribution 
which is proportional to a singularity
$\delta^{(4)}(0)$, and dimensional regularization sets it equal to zero.
(A similar remark was in order when a change of field variables
allowed to cross out $W_4(X)$ from the lagrangian in (\ref{3.7})).

We find,
\begin{eqnarray}
& &\frac{1}{2}\sigma^{\alpha \beta} \sigma^{\beta \alpha} =
\left( \frac {1}{8} + \frac {n_l^2}{32} \left( 
\frac {W_1'}{W_1} \right)^4 \right)
\langle C_\mu C_\nu \rangle \langle C^\mu C^\nu \rangle +
\frac {1}{16}\langle C_\mu C^\mu \rangle \langle C_\nu C^\nu \rangle 
-\frac {1}{4} \langle C_\mu C^\mu C_\nu \rangle \langle C^\nu \rangle
\nonumber \\
&+& \frac {n_l}{8} \left( \frac {W_1'}{W_1} \right)^2 \left( 
\langle C_\mu C_\nu C^\mu C^\nu \rangle -
\langle C_\mu C^\mu C_\nu C^\nu \rangle \right) +
\frac {1}{16} \langle R \rangle^2 + \frac{n_l}{16} \langle R^2 \rangle
+ \frac {n_l}{2} S^2 +
\frac {1}{4} \langle T \rangle^2  \nonumber \\
&+&\frac {n_l}{4} \langle T^2 \rangle
+ S \langle T \rangle
+\frac {n_l}{8} \left(\left( \frac {W_1'}{W_1}\right)^2 -1 \right)
\langle C_\mu C^\mu R \rangle
- \frac {1}{8} \langle C_\mu C^\mu\rangle \langle R \rangle +
\frac {1}{4} \langle C_\mu \rangle \langle  C^\mu R \rangle \nonumber\\
&+& \frac {1}{2} S \langle C_\mu \rangle \langle C^\mu \rangle +
\frac {n_l}{2} \left( \frac {1}{2} \left( \frac {W_1'}{W_1} \right)^2 
-1 \right)
S \langle C_\mu C^\mu \rangle +
\frac {n_l}{2} S \langle R \rangle + \frac {1}{2} \langle R T \rangle
+ \frac {\nz}{4} \left( \frac {W_1'}{W_1}\right)^2 
 \langle C_\mu \rangle \langle  C^\mu T \rangle,
\nonumber
\end{eqnarray}
and
\begin{eqnarray}
& &\frac {1}{12} R_{\mu \nu}^{(\alpha \beta)}R^{\mu \nu \ (\beta \alpha)}=
- \frac {n_l}{24} \langle Q_{\mu \nu} Q^{\mu \nu} \rangle 
+\frac {1}{24} \langle Q_{\mu \nu} \rangle \langle  Q^{\mu \nu} \rangle
+\frac {i \nz}{24} \left( \frac {W_1'}{W_1} \right)^2 
\langle Q_{\mu \nu}[C^\mu,C^\nu] \rangle  \nonumber \\
&+& \frac{1}{24}\langle H_{\mu \nu} H^{\mu \nu} \rangle 
- \frac {n_l}{24}  \langle H_{\mu \nu} \rangle \langle  H^{\mu \nu} \rangle
+\frac {n_l^2}{96} \left( \frac {W_1'}{W_1} \right)^4
\left(\langle C_\mu C_\nu \rangle 
\langle C^\mu C^\nu \rangle - \langle C_\mu C^\mu \rangle 
\langle C_\nu C^\nu \rangle \right)  \nonumber \\
&+& \frac {n_l}{12} \left( \frac {W_1'}{W_1} \right)^2 \langle C_\mu
H^{\mu \nu} \rangle \langle C_\nu \rangle, \nonumber
\end{eqnarray}
which are the only structures that get divergent contributions at one-loop
(see Appendix A).
 
Since quantum scalar fields in four dimensions do not generate
any anomaly to the $U_L \otimes U_R$ symmetry, the new 
terms needed to renormalize the one-loop result are necessarily in the list
of all possible operators of $O(p^4)$ invariant under $U_L \otimes U_R$. 

The list of independent operators is given below.
The criteria used to select this particular set
are the following: terms involving the derivatives
$D_\mu M$, $D_\mu N$, $D_\mu F_L^{\mu \nu}$ and alike;
three derivatives of $\hat{\theta}$ 
 $(D^\mu D_\mu D_\nu \hat{\theta})$, $(D^\mu D_\nu D_\mu \hat{\theta})$;
$D^\mu D_\mu C_\nu$,
$(D_\mu D_\nu \hat{\theta})$ or $D_\mu C_\nu$ can be removed as 
combinations of those in (\ref{3.18}) plus terms with the piece
$D_\mu C^\mu$: finally,
these can be eliminated with the equations of motion (\ref{3.10}).
The two derivatives of $\hat{\theta}$ can always
be chosen to appear under the form $D^\mu D_\mu \hat{\theta}$.
The rest of operators that are not independent
have been removed upon integration by parts and with the help of
the identities
$$D^\mu C^\nu - D^\nu C^\mu = -[C^\mu, C^\nu]+ i F_L^{\mu \nu} -
i U^\dagger F_R^{\mu \nu} U , $$
$$[D^{\mu}, D^{\nu}] C^\rho = -i[F_L^{\mu \nu}, C^\rho],$$
$$[D^{\mu}, D^{\nu}] \hat{\theta} = i\langle  F_R^{\mu \nu}- 
F_L^{\mu \nu}\rangle .$$
All the following terms are real $O_i^\dagger=O_i$.
The first ones correspond to the twelve $O(p^4)$ operators
of $SU_L \otimes SU_R$
(recall that $C_\mu \equiv U^\dagger D_\mu U$, 
$M \equiv U^\dagger \chi + \chi^\dagger U$,
$N \equiv U^\dagger \chi - \chi^\dagger U$ and $\hat{\theta}= i \theta$),
\begin{eqnarray}
O_0 &=& \langle  D_\mu U \ D_\nu U^\dagger \ D^\mu U \ D^\nu U^\dagger \rangle 
=\langle C^\mu C^\nu C_\mu C_\nu\rangle ,\nonumber \\
O_1 &=& \langle  D_\mu U^\dagger \ D^\mu U\rangle ^2 =\langle C^\mu C_\mu\rangle \langle C^\nu C_\nu\rangle 
\nonumber, \\
O_2 &=& \langle  D_\mu U^\dagger \ D_\nu U \rangle \langle  D^\mu U^\dagger \ 
D^\nu U\rangle  
=\langle C^\mu C^\nu\rangle \langle C_\mu C_\nu\rangle ,\nonumber \\
O_3 &=& \langle  D_\mu U^\dagger \ D^\mu U \ D_\nu U^\dagger \ D^\nu U \rangle  
=\langle C^\mu C_\mu C^\nu C_\nu\rangle  ,\nonumber \\
O_4 &=& \langle  D_\mu U^\dagger \ D^\mu U \rangle  \langle U^\dagger \chi +
 \chi^\dagger U \rangle 
=- \langle C^\mu C_\mu\rangle \langle M\rangle , \nonumber \\
O_5 &=&\langle  D_\mu U^\dagger \ D^\mu U \  
( U^\dagger \chi + \chi^\dagger U ) \rangle 
= - \langle C^\mu C_\mu M\rangle , \nonumber \\
O_6 &=& \langle U^\dagger \chi + \chi^\dagger U \rangle ^2 = \langle M\rangle ^2, \nonumber \\
O_7 &=& \langle U^\dagger \chi - \chi^\dagger U \rangle ^2 = \langle N\rangle ^2, \nonumber \\
O_8 &=& \langle  \chi^\dagger U \chi^\dagger  U +
 U^\dagger \chi U^\dagger \chi \rangle 
= \frac {1}{2}  \langle M^2  +  N^2\rangle
, \nonumber \\
O_9 &=& -i \langle  F^{\mu \nu}_R \ D_\mu U \ D_\nu U^\dagger +
F^{\mu \nu}_L \ D_\mu U^\dagger \ D_\nu U \rangle  = i \langle  C_\mu C_\nu
\left( F_L^{\mu \nu} + U^\dagger F_R^{\mu \nu} U \right)\rangle ,
\nonumber \\
O_{10} &=& \langle  U^\dagger F_R^{\mu \nu} U F_{L\; \mu \nu}\rangle , \nonumber \\
O_{11} &=& \langle  F_{R \; \mu \nu} F_R^{\mu \nu} + F_{L \; \mu \nu} F_L^{\mu \nu}\rangle ,
\nonumber \\
O_{12} &=& \langle  \chi^\dagger \chi \rangle  =\frac {1}{4} 
 \langle M^2 - N^2 \rangle  ,  \nonumber
\end{eqnarray}

The following eight operators are obtained from the previous twelve by
splitting up single traces into products of traces:  
$\langle C_\mu\rangle $ does not vanish, 
$\langle C_\mu\rangle  \neq 0$, for $U_L \otimes U_R$.
They read,
\begin{eqnarray}
O_{13} &=& \langle  U^\dagger \ D_\mu U\rangle \langle U^\dagger \ D^\mu U \ 
D_\nu U^\dagger
\ D^\nu U\rangle =-\langle C^\mu\rangle \langle C_\mu C^\nu C_\nu\rangle ,
\nonumber \\
O_{14} &=& \langle  U^\dagger \ D_\mu U\rangle \langle  U^\dagger \ D^\mu U\rangle 
\langle D^\nu U^\dagger \ D_\nu U\rangle =-\langle C^\mu\rangle \langle C_\mu\rangle \langle C^\nu C_\nu\rangle  , \nonumber \\
O_{15} &=& \langle  U^\dagger \ D_\mu U\rangle \langle  U^\dagger \ D_\nu U\rangle 
\langle D^\mu U^\dagger \ D^\nu U\rangle =-\langle C^\mu\rangle \langle C^\nu\rangle \langle C_\mu C_\nu\rangle  ,\nonumber \\
O_{16} &=& \langle  U^\dagger \ D^\mu U\rangle \langle  U^\dagger \ D_\mu U\rangle 
\langle  U^\dagger \ D^\nu U\rangle \langle  U^\dagger \ D_\nu U\rangle =
 \langle C^\mu\rangle \langle C_\mu\rangle \langle C^\nu\rangle \langle C_\nu\rangle  ,\nonumber\\
O_{17} &=& \langle  U^\dagger \ D^\mu U\rangle \langle  U^\dagger \ D_\mu U\rangle 
\langle U^\dagger \chi + \chi^\dagger U \rangle =\langle C^\mu\rangle \langle C_\mu\rangle \langle M\rangle  , \nonumber \\
O_{18} &=& \langle  U^\dagger \ D_\mu U\rangle  \langle  D^\mu U^\dagger\ \chi
 -D^\mu U\ 
\chi^\dagger \rangle =- \langle C^\mu\rangle \langle C_\mu M\rangle , \nonumber \\
O_{19} &=&  \langle  F_{R \; \mu \nu}\rangle  \langle F_R^{\mu \nu}
\rangle  +
\langle  F_{L \; \mu \nu}\rangle  \langle F_L^{\mu \nu}\rangle  ,
\nonumber \\
O_{20} &=& \langle F_{R \; \mu \nu}\rangle \langle F_L^{\mu \nu}\rangle  
.\nonumber
\end{eqnarray}

So far, all are {\it parity even} operators.
The next seven are similar but have {\it odd parity},
\begin{eqnarray}
O_{21} &=& -i\langle  D_\mu U^\dagger \ D^\mu U \; ( U^\dagger \chi - 
\chi^\dagger U) \rangle  = i \langle  N C^\mu C_\mu\rangle  ,\nonumber\\
O_{22} &=& -i \langle  D_\mu U^\dagger \ D^\mu U) \rangle  \langle U^\dagger \chi - 
\chi^\dagger U \rangle  = i \langle  N \rangle \langle C^\mu C_\mu\rangle  ,\nonumber\\
O_{23} &=& i\langle  U^\dagger \ D_\mu U\rangle  \langle  D^\mu U^\dagger\
 \chi -D^\mu U\ 
\chi^\dagger \rangle  = i \langle  N C^\mu\rangle \langle  C_\mu\rangle  ,\nonumber\\
O_{24} &=& i \langle  U^\dagger \ D^\mu U\rangle \langle  U^\dagger \ D_\mu U\rangle 
\langle U^\dagger \chi - \chi^\dagger U \rangle = i \langle  N \rangle \langle C^\mu\rangle \langle C_\mu\rangle  ,\nonumber\\
O_{25} &=& 
i \langle U^\dagger \chi U^\dagger \chi - \chi^\dagger U \chi^\dagger  U \rangle  =
i \langle NM\rangle  ,\nonumber\\
O_{26} &=& i \left( \langle U^\dagger \chi\rangle ^2 - \langle  \chi^\dagger U\rangle ^2 \right) =
i \langle N\rangle \langle M\rangle  ,\nonumber\\
O_{27} &=&  \langle  U^\dagger \ D_\mu U\rangle 
\langle F_L^{\mu \nu} U^\dagger \ D_\nu U-F_R^{\mu \nu}
 \ D_\nu U\ U^\dagger\rangle = 
\langle C_\mu\rangle \langle C_\nu \left( F_L^{\mu \nu} - 
U^\dagger F_R^{\mu \nu}U\right)\rangle  ,\nonumber\\
\nonumber
\end{eqnarray}

Three operators involve the $\epsilon_{\mu \nu \rho \sigma }$ tensor.
The first two of them are $odd$ under parity and $O_{30}$ is $even$.
\begin{eqnarray}
O_{28} &=& \epsilon_{\mu \nu \rho \sigma}
\langle F_L^{\mu \nu} U^\dagger F_R^{\rho \sigma} U \rangle ,\nonumber \\
O_{29} &=&  i \epsilon_{\mu \nu \rho \sigma} 
\langle  \left( F_L^{\mu \nu} + 
U^\dagger F_R^{\mu \nu} U \right) C^\rho C^\sigma\rangle ,\nonumber \\
O_{30} &=& \epsilon_{\mu \nu \rho \sigma} \langle  \left( F_L^{\mu \nu} - 
U^\dagger F_R^{\mu \nu} U \right) C^\rho\rangle \langle C^\sigma\rangle
,\nonumber \\
\label{3.18}
\end{eqnarray}

From $O_{31}$ to $O_{57}$ they involve derivatives
of the source $\hat{\theta}$, and are given in Appendix C.

The operators that appear at $O(p^2)$ in (\ref{3.8}) are
\begin{equation}
\begin{array}{ll}
E_1 = -\langle C^\mu C_\mu\rangle  &\hskip 2cm
E_4 = \langle C^\mu\rangle \langle C_\mu\rangle \cr 
E_2 = \langle M\rangle  &\hskip 2cm 
E_5 = \langle C^\mu\rangle  D_\mu \hat{\theta}\cr
E_3 = i\langle N\rangle  &\hskip 2cm
E_6 = D^\mu\hat{\theta} D_\mu \hat{\theta}
\end{array}
\label{3.22p}
\end{equation}

The effective lagrangian including up to one-loop corrections is, thus,
\begin{equation}
{\cal L}^{One-loop}=-W_0^r(X,\mu)+ \sum_{i=1}^{6} W_i^r(X,\mu) E_i +
\sum_{i=0}^{57} \beta_i^r(X,\mu) O_i + \; finite \;\; non-local .
\label{3.22}
\end{equation}

The following structure of counter-terms
renders it finite (see Appendix A):
\begin{eqnarray}
\delta W_i (X) &=&  \hbar W_i^{(1)} (X,\mu) + \hbar 
\Omega_i(X,\mu) \lambda_\infty  + O(\hbar^2), \;\; i=0,...,6 \;, \nonumber \\
\beta_i (X) &=&\beta_i^r (X,\mu) + \hbar B_i(X,\mu) \lambda_\infty 
+ O(\hbar^2), \;\; i=0,...,57. 
\label{3.19}
\end{eqnarray}
with
\begin{equation}
\lambda_\infty=\frac {\mu^{D-4}}{(4 \pi)^2} \left(\frac {1}{D-4} -
\frac{1}{2} \left( \log 4\pi - \gamma +1 \right) \right),
\label{3.20}
\end{equation}
so that
\begin{equation}
W_i^r (X,\mu)= W_i (X) + \hbar W_i^{(1)} (X,\mu), \;\; i=0,...,6 \;.
\label{3.21}
\end{equation}
At this point we have reinserted the powers of $\hbar$ to help the counting
of loops. Recall that the one-loop effective action carries one
power of $\hbar$. 

The roster of functions $\Omega_i$'s and $B_i$'s is given in Appendix D;
they are the main result of our paper.
The counter-terms in (\ref{3.19}) are written in terms of the functions
$\Omega_i$'s and $B_i$'s for the sake of concision. This notation, however,
may seem a bit contrived. It should be read in the usual way of perturbation
theory, namely with the functions $\Omega_i$'s and $B_i$'s 
understood as their series in powers of $X$. The
renormalization of the functions means the renormalization of the coupling
constants which are the coefficients in these expansions.

Parity, charge conjugation and time reversal ought to be conserved.
Only the operators that are invariant under charge conjugation themselves
can appear in the lagrangian. This is because $X$ is invariant under $C$ 
as well (see Appendix E). The list of $C$-violating operators is given
in Appendix C.

The result is valid for any value of $n_l$.
However, depending on the specific $n_l$ considered,
there are $n_l$-dependent factorization relations among the
traces of products of $n_l \times n_l$ matrices that are of relevance to
us since some operators in the list become redundant, i.e., some can be written
in terms of a smaller subset.
These relations follow from the Cayley-Hamilton
theorem and have been extensively used in \cite{gl}. For $n_l=3$
they boil down to \cite{fearingscherer}
$$ A^3 - \langle A\rangle A^2 + \frac {1}{2} \left(
\langle A\rangle ^2 -\langle A^2\rangle  \right) A - \det (A)=0,$$
for any $3 \times 3$ matrix, be it hermitian or not, and to
$$\sum_{6 \; perm} \langle A_1 A_2 A_3 A_4\rangle  - \sum_{8 \; perm}
\langle A_1 A_2 A_3\rangle \langle A_4\rangle 
-\sum_{3 \; perm} \langle A_1 A_2\rangle \langle A_3 A_4\rangle   $$
\begin{equation}
+\sum_{6 \; perm} \langle A_1 A_2\rangle \langle A_3\rangle \langle A_4\rangle 
-\langle A_1\rangle \langle A_2\rangle \langle A_3\rangle 
\langle A_4\rangle =0.
\label{3.23}
\end{equation}
The first relation ensures that the determinant of a matrix can always
be expressed in terms of traces and justifies why the determinants of products
of $U$ matrices and their derivatives need not be considered independently.
From the second relation, with
\begin{eqnarray}
A_1=A_2=C_\mu, &\;\;\;\;\;\;\;\;\;\;& 
A_3=A_4=C_\nu,
\label{3.24}
\end{eqnarray}
and summing over the indices $\mu$ and $\nu$, one gets
\begin{equation}
2 O_0 - O_1 -2 O_2 + 4 O_3 + 8 O_{13} - 2 O_{14} - 4 O_{15} - O_{16}=0.
\label{3.25}
\end{equation}
Thus, for $n_l=3$ we can spare $O_0$ and
the $O(p^4)$ lagrangian will be written as
\begin{equation}
{\cal L}_4=\sum_{i=1}^{57} L_i (X) O_i,
\label{3.26}
\end{equation}
where
\begin{equation}
\begin{array}{ll}
L_1 = \beta_1 + \frac {1}{2} \beta_0 &\hskip 1cm 
L_{13} = \beta_{13} - 4 \beta_0\cr
L_2 = \beta_2 + \beta_0 &\hskip 1cm 
L_{14} = \beta_{14} + \beta_0\cr
L_3 = \beta_3 - 2 \beta_0 &\hskip 1cm
L_{15} = \beta_{15} + 2 \beta_0\cr 
&\hskip 1cm L_{16} = \beta_{16} + \frac {1}{2} \beta_0 
\end{array}
\label{3.27}
\end{equation}
and $L_i = \beta_i$ for the rest.


From $O_1$ to $O_9$ the same notation as in 
$SU_L \times SU_R$ \cite{gl} has been kept, also for the coefficient 
functions.  
The constants $H_1$, $H_2$ in \cite{gl} have turned into the functions 
$L_{11}(X)$, $L_{12}(X)$. The new operators that 
appear in this $U_L \otimes U_R$ lagrangian are labeled
from $O_{13}$ onwards and the coefficient functions $L_i(X)$'s follow suit.

\vspace*{0.5cm}
If one disregards all the coefficients associated to $U_A(1)$ one finds
for the $B_i$'s in (\ref{3.19})
\begin{equation}
\begin{array}{lllllll}
B_0=\frac {n_l}{48} & B_1= \frac {1}{16} & B_2 = \frac {1}{8} &
B_3=\frac {n_l}{24} & B_4= \frac {1}{8} & B_5 = \frac {n_l}{8} &
B_6= \frac {1}{16} \\
B_7 = 0 & B_8= \frac {n_l}{16} & B_9=\frac {n_l}{12} & B_{10}=-\frac {n_l}{12}&
B_{11}=-\frac {n_l}{24} & B_{12}= \frac {n_l}{8} & B_{13}= \frac {1}{4} \\
B_{14}=0 & B_{15}=0 & B_{16} = 0 & B_{17}=0 & B_{18}= - \frac {1}{4} & 
B_{19}= \frac {1}{24} & B_{20}= \frac {1}{12} \\
\end{array}
\label{llista}
\end{equation}
$O_{30}$ involves an
$\epsilon_{\mu \nu \rho \sigma}$ and does not need to be renormalized.

In the case of $n_l=3$, taking into account that the same relations
from (\ref{3.27}) should hold, and renaming the constants as 
$\Gamma_i$'s, we obtain
\begin{equation}
\begin{array}{lllllll}
\Gamma_1  = \frac {3 } {32 } &
\Gamma_2  = \frac {3 } {16 } &
\Gamma_3  = 0 &
\Gamma_4  = \frac {1 } {8 } &
\Gamma_5  = \frac {3 } {8 } &
\Gamma_6  = \frac {1 } {16 } &
\Gamma_7  = 0 \\
\Gamma_8  = \frac {3 } {16 } &
\Gamma_9  = \frac {1 } {4 } &
\Gamma_{10}  = - \frac {1 } {4 } &
\Gamma_{11}  = - \frac {1 } {8 } &
\Gamma_{12}  = \frac {3 } {8 } &
\Gamma_{13}  = 0 &
\Gamma_{14}  = \frac {1 } {16 } \\
\Gamma_{15}  = \frac {1 } {8 } &
\Gamma_{16}  = \frac {1 } {32 } &
\Gamma_{17}  = 0 &
\Gamma_{18}  = - \frac {1 } {4 } &
\Gamma_{19}  = \frac {1 } {24 } &
\Gamma_{20}  = \frac {1 } {12 }. &  
\end{array}
\nonumber
\end{equation}
They coincide to those of $SU_L(3) \otimes SU_R(3)$
\cite{gl} except for the terms that involve
$\langle M^2 \rangle$, $\langle M \rangle^2$,
$\langle N^2 \rangle$, $\langle N \rangle^2$: $\Gamma_6$, $\Gamma_8$,
$\Gamma_{12}$. The reason is that among the building blocks that have been
used to write the chiral lagrangian, $C_\mu$ and $F_{L, R}^{\mu \nu}$
have vanishing traces in the case of
$SU_L(3) \otimes SU_R(3)$, whereas neither $M$ nor $N$ do.
Although it is less immediate
to retrieve the $SU_L(3) \otimes SU_R(3)$ coefficients
\begin{equation}
\begin{array}{lll}
\Gamma_6^{[SU]}= \frac {11}{144} & \Gamma_8^{[SU]}= \frac {5}{48} &
\Gamma_{12}^{[SU]}= \frac {5}{24},
\end{array}
\nonumber
\end{equation}
from our result, there are simple relations that have
to be verified. For instance, if one adds all the divergent pieces
that go with the operators $O_6$, $O_8$, $O_{12}$,
sets $\chi=m^2 I$ for simplicity,
and expands the operators, it is easy to check that the 
$SU_L(3) \otimes SU_R(3)$ and the $U_L(3) \otimes U_R(3)$ coefficients of
$\langle \chi^\dagger \chi \rangle$ 
and $\vec{\pi}^2$ verify, respectively,
$$\frac {9}{8} \left(12 \Gamma_6^{[SU]} + 2 \Gamma_8^{[SU]} 
+ \Gamma_{12}^{[SU]} \right)=
\left( 12 \Gamma_6 + 2 \Gamma_8  + \Gamma_{12} \right),$$
and similarly
$$\frac {9}{8} \left( 3 \Gamma_6^{[SU]} + \Gamma_8^{[SU]} \right)=
\left( 3 \Gamma_6 + \Gamma_8 \right).$$
One recognizes the ratio $\frac {9}{8}$ as the fraction
of degrees of freedom in the two theories, for these coefficients
multiply one-loop divergent pieces which in the two cases stem from
tadpole diagrams, which give a constant divergent contribution 
for all the virtual mesons that travel around the loop.  
Therefore, the total result is
proportional to the number of degrees of freedom that in each case
can circulate.
(Of course the same argument goes through for any number of flavours
$n_l$ and a similar result holds in general.
The $SU_L(n_l) \otimes SU_R(n_l)$ coefficients are \cite{donoghue}
$\Gamma_6^{[SU]} = \frac {2 + n_l^2}{16 n_l^2}$,
$\Gamma_8^{[SU]} = \frac {n_l^2-4}{16 n_l}$. The second relation holds in the
form $\frac {n_l \Gamma_6 + \Gamma_8}{n_l^2}= 
\frac {n_l \Gamma_6^{[SU]} + \Gamma_8^{[SU]}}{n_l^2-1}$. The first 
relation, that
now involves the combination $4n_l \Gamma_6 + 2 \Gamma_8 + \Gamma_{12}$,
reduces to the previous one if one realizes that $\Gamma_{12}=2 \Gamma_8$
in either case.)

\vspace*{0.5cm}
One important difference between the $SU_L \otimes SU_R$ case and
ours is that in the first theory the meson masses do not get any 
infinite contribution and in this case they do.
This statement needs some qualification for the language it uses is
the customary of renormalizable field theories, where
the divergences that are generated require a fixed number of counter-terms, 
of same type as the terms in the lagrangian only.
The chiral expansion in increasing number
of derivatives is not a theory of this kind, rather it is non-renormalizable,
because at each higher loop new terms are required to absorb the new infinities.
In $SU_L \otimes SU_R$ one only needs
counter-terms of a chiral order higher
than the terms involved in the loops. In $U_L \otimes U_R$ 
we find a combination of both previous cases.
It is non-renormalizable and there is a $O(p^0)$ term (\ref{3.7}), 
included to reproduce the $U_A(1)$ anomaly,
which at one-loop induces
a mixture of chiral orders in the divergent parts, as can be seen 
from the heat-kernel expressions: the
divergences are proportional to $\sigma^2$ (\ref{a.4}) and
$\sigma= \sigma_{(0)} + \sigma_{(2)}$ decomposes in (\ref{3.15}) in two 
pieces, $O(p^0)$ and $O(p^2)$, respectively. (There are divergences 
proportional to $R^2$ too in (\ref{a.4}), but
the curvature $R$ associated
to the connexion (\ref{3.14p}) does not get any $O(p^0)$ contribution).

There is a lot of freedom in deciding the prescriptions of what
removes which divergences, all of them equally acceptable from the
point of view of rendering the final result finite.
They are not completely arbitrary, though, since the nesting of divergences
when higher loops are considered imposes some constraints among the results
at different orders, of the Gell-Mann and Low type in QED \cite{weinberg}.
Furthermore, some of them
appear more natural than others.

Let us analyse the question of the renormalization of
the pseudo-goldstone boson
masses, induced by a quark-mass term. Let us first
disregard the $U_A(1)$ anomaly
by freezing the functions of $X$ to their constant values at $X=0$
and, for simplicity,
consider the symmetric case
where the three quark species are degenerate in mass,
and switch the external sources off: $\chi$ is a constant
that multiplies the unit matrix; at tree-level $\chi$ 
gives the nonet mass. Now, let us write all the terms 
quadratic in the fields with at most two derivatives, 
having added the one-loop
divergences to the tree-level result (prior to renormalization). There are
contributions from the operators $E_1$, $E_2$ in (\ref{3.22p}) 
for the tree-level parts, whereas the 
divergent parts come with the operators $O_4$, $O_5$, $O_6$,
$O_8$, $O_{17}$ and $O_{18}$, and can be read off from (\ref{3.19}) 
and (\ref{llista}).
It yields,
\begin{equation}
\left(1 - 2 n_l \frac {\chi}{f^2} \hbar \lambda_\infty \right)
\left( \frac {1}{2} \partial_\mu \vec{\pi} \cdot \partial^\mu \vec{\pi}
- \frac {1}{2} \chi \vec{\pi} \cdot \vec{\pi} \right) +
\frac {1}{2} \partial_\mu \eta^0 \partial^\mu \eta^0 
- \frac {1}{2} \left(1 - 2 n_l \frac {\chi}{f^2} \hbar \lambda_\infty \right)
\chi (\eta^0)^2, 
\label{ren}
\end{equation}
where $\lambda_\infty$ is the ultraviolet divergent amount, 
that in dimensional
regularization is essentially $\frac {1}{D-4}$, (see \ref{3.20}).
The difference between the singlet and the octet is apparent. The
piece $\partial_\mu \eta^0 \partial^\mu \eta^0$
does not get any divergent contribution while the octet
counterpart $\partial_\mu \vec{\pi} \cdot \partial^\mu \vec{\pi}$
does,  and exactly by the same amount as the mass term
$\vec{\pi} \cdot \vec{\pi}$ does as well. In the octet sector,
one can pull out the common factor from the entire kinetic term and 
render a finite result by a field redefinition 
$\pi \to (1-\hbar n_l \frac {\chi}{f^2} \lambda_{\infty} ) \pi$.
There is no infinity left over that could require mass renormalization.

In chiral perturbation theory it is often simpler
to talk about a renormalization of an $O(p^4)$ operator
rather than a wave-function renormalization, 
but in our case this is what it corresponds to,
and this precision is required to qualify the mass renormalization issue.
The same result also holds in $SU_L \otimes SU_R$.

For the $\eta^0$, though, the divergent contributions to $\partial_\mu
\eta^0 \partial^\mu \eta^0$ - which are none -, and to $(\eta^0)^2$
are different, and therefore
the mass gets necessarily renormalized by an infinite amount.
This is a remarkable difference between
$SU_L \otimes SU_R$ and $U_L \otimes U_R$.

In both cases, of course,
the divergences disappear if the quark-mass is turned off
$\chi \to 0$, which reflects the fact that the spontaneously
broken symmetry is built-in, loop by loop, in the quantum theory
and prevents the goldstone particles
from acquiring a mass. When the quark-mass is turned on,
it is not true that the symmetry structure prevents 
the $\eta^0$ mass from being infinitely renormalized, as happens
with the octet mass.
The difference can be rooted to the terms in the
lagrangian that are responsible for the wave-function renormalization. 
To one-loop, this only includes the
terms that are quartic in the fields with two derivatives, which are obtained
by expanding the operator $\langle C^\mu C_\mu \rangle$; by contracting
the two fields that carry no derivatives a tadpole diagram is generated
and its divergence multiplies 
$\partial_\mu \vec{\pi} \cdot \partial^\mu \vec{\pi}$. 
The interesting point is that in the chiral lagrangian such
a term involving the $\eta^0$ field, like
$(\eta^0)^2 \partial_\mu \eta^0 \partial^\mu \eta^0$ 
or $\vec{\pi} \cdot \vec{\pi} \partial_\mu \eta^0 \partial^\mu \eta^0$, 
does not exist at all, as can be
immediately seen by making the invariant decomposition of
the matrix $U$ in $U=e^{i \sqrt{\frac {2}{n_l}} \frac {\eta^0}{f}} U_s$,
where $U_s$ contains only the octet fields and has $\det U_s=1$. $C_\mu$
then reads
$$C_\mu \equiv U^\dagger \partial_\mu U = i \sqrt{ \frac {2}{n_l}}
\frac {1}{f} \partial_\mu \eta^0 + U_s^\dagger \partial_\mu U_s,$$
and $\langle C^\mu C_\mu \rangle$
\begin{equation}
\langle C^\mu C_\mu \rangle = -\frac{2}{f^2} \partial^\mu \eta^0 \partial_\mu \eta^0
+\langle U_s^\dagger \partial_\mu U_s \ U_s^\dagger \partial^\mu U_s \rangle,
\end{equation}
and no crossed singlet-octet term survive for 
$\langle U_s^\dagger \partial_\mu U_s \rangle =0$. 
$\langle C^\mu C_\mu \rangle$ provides the 
$\eta^0$ with a kinetic term and nothing else, and no term can generate
a one-loop wave-function divergence for it.
Whereas for the octet, one learns from 
$$U_s^\dagger \partial_\mu U_s= i \ \frac {1}{f} \partial_\mu \pi
+ \frac {1}{2f^2} [\pi,\partial_\mu \pi] - i  \ \frac {1}{6} \frac {1}{f^3}
\left[ \pi, [\pi, \partial_\mu \pi] \right] + \; ...$$
that 
\begin{equation}
- \frac {f^2}{4} \langle U_s^\dagger 
\partial^\mu U_s \ U_s^\dagger \partial_\mu U_s \rangle
= \frac {1}{2} \partial_\mu \vec{\pi} \cdot \partial^\mu \vec{\pi}+
\frac {1}{48f^2} \langle[\pi,\partial_\mu \pi]
[\pi,\partial^\mu \pi] \rangle + \; ... \  .
\label{qtermes}
\end{equation}
It is the second term in (\ref{qtermes})
which is responsible for the one-loop wave-function renormalization.
The structure of commutators makes one realize again that 
such terms vanish for the singlet.

This same piece generates also a divergence to the mass term. It gets
additional $O(p^2)$ contributions from 
operator $E_2$ in (\ref{3.22p}) which come from tadpole diagrams too
that arise from four meson interaction terms
$\sim \frac {\chi}{f^2} \langle \Phi^4 \rangle$.
It is a well-know result that a lagrangian for scalar fields
with no derivative couplings
other than the kinetic energy with a quartic interaction
has an effective action that needs a wave-function renormalization
that starts at two loops \cite{iim}. At one-loop it requires
mass renormalization and this is what we find for the $\eta^0$ field. 
(Vertices with more
than four fields from $E_2$ at one-loop do not participate in
the renormalization of the kinetic terms).


We see that the singlet vs. octet difference in mass renormalization 
is imbued by the structure of the symmetry group.

If one includes the $U_A(1)$-anomaly effects, both the octet and 
the singlet get their masses renormalized by an infinite
amount.



\vspace*{0.5cm}
In this section we have presented the complete one-loop calculation
of the divergent part of the effective action. It is a calculation to all 
orders in $1/N_c $, in the sense that the functions of $X$ 
have been kept generic through the end. 
There are many unknown parameters in this approach (potentially,
all the coefficients of the functions of $X$) and without any further
restrictions we would not know how to eliminate any of them.
We invoke the $1/N_c$ expansion of QCD 
and the restrictions it imposes on the chiral lagrangian to classify 
the coefficients
according to the maximum $1/N_c$ power allowed for each,
so as to estimate their size, 
and with this criterion try to select
the fewer terms that allegedly bring the main contributions.
This will be done in the next section.

\section{The $1/N_c$ expansion.}
The systematic expansion of QCD in powers of $1/N_c$ provides
a way of effectively reducing the 
number of constants that intervene in a certain process, once 
it is decided where to truncate the series in $1/N_c$. 
If $N_c$ is large enough, a few terms will suffice to give a
good account of the exact result. How much large is
{\it large enough} is a question hard to assess,
for a good reason, that despite the
simplification the large $N_c$ limit entails
technically, it remains too difficult to sum 
the subclass of diagrams that survive in the limit 
and it worsens,
if anything, for the sub-leading contributions. 
It is argued that, conceivably,
big numerical factors might be 
accompanying the powers of $N_c$ in the denominator; in that
case a few terms would give accurate predictions even for $N_c=3$. 
That would explain the remarkable resemblance of many qualitative
features and patterns of the leading terms
to those observed in hadron physics, with $N_c=3$ 
\cite{Witten}. At any rate, lacking of any analytical result,
it is the accuracy of the predictions
in explaining the data what could give 
an ultimate justification for the expansion. It is this perspective 
what launched
this project, to set out the basis for the systematic
study of the predictions that come out from such scenario
for soft $\pi$, $K$, $\eta$, $\eta'$ 
so as to discern in what processes and to what extent an 
agreement with experiment holds.

On general grounds, Witten \cite{Witten} showed that in the 
$N_c \to \infty$ limit, if QCD confines it has a mesonic spectrum 
that consists of an infinite number of noninteracting, stable states,
with masses that have smooth and finite limits. Furthermore, 
the strong interaction scattering amplitudes are given, to
lowest order in $1/N_c$, by sums of tree diagrams with mesons
exchanged, which can be derived from an effective lagrangian with 
local vertices and local meson fields. 
The decay constants $f$'s are of order
$\sqrt{N_c}$  and a coupling constant for a vertex with $k$ mesons
attached to it is of order $ {N_c}^{-\frac {(k-2)}{2}}$, i.e., it decreases
with the multiplicity of mesons in the vertex, each new meson bringing
in one additional power of $1/\sqrt{N_c}$.
The $N_c$ counting rules imply that while large-$N_c$ QCD is a
strong interacting theory in terms of quarks and gluons, it is equivalent
to a weakly interacting theory of mesons. 
The higher order corrections in the $1/N_c$ expansion,
in addition to new couplings in the effective theory, also include
the loop diagrams of mesons, which as in any quantum field theory 
reestablish the unitarity constraints
on the amplitudes (cuts, discontinuities across, etc...).
Each loop of mesons, in the effective theory, contributes an extra power 
of $1/N_c$.

In addition to the mesons there are infinitely many glueball states, which at
$N_c \to \infty$ are stable and noninteracting. 
The amplitude for a glueball
to mix with a meson is of order $O(1/\sqrt{N_c})$, and the vertices
in which they are involved are even more suppressed than the meson vertices:
one power of $1/N_c$ for each glueball.

However, only $\pi$, $K$, $\eta$, $\eta'$ remain massless in the
chiral limit and $N_c \to \infty$, their masses being precluded by
Goldstone's theorem. The rest of excited mesons and glueballs
remain massive, with
a typical hadronic mass of about 1 GeV. They decouple from the
soft processes that involve the goldstone bosons by virtue of their masses
and they are integrated out from the effective theory. The baryons 
have masses much higher for large $N_c$, for they are known to grow
like $N_c$ \cite{Witten}.

Within the large-$N_c$, the $U_A(1)$ anomaly effects can be accommodated
in a rather natural way in the framework of the chiral lagrangian. 
Actually, the identification of each ingredient that has been taken into
account in writing down the effective theory is clearcut:
the constraints
imposed on the interactions among goldstone bosons are contained in the
operators that involve the $U$ matrices (but not $\log U$);
the quark masses enter through
the terms in $\chi$; and the $U_A(1)$ enters through the functions of $X$.
One can talk of switching off the $U_A(1)$ anomaly by freezing the functions
to constants, in a similar way as one can take the chiral limit by sending
the quark masses to zero. In this language, one can say that 
the $\eta^0$ has two ways of manifesting itself: either as a goldstone boson
or as the argument of the functions of $X$, breaking chiral
symmetry as dictated by the $U_A(1)$ anomaly.
When it manifests as a goldstone boson, its couplings 
follow the rules of the meson couplings.
This is its the mesonic part, associated to the content in quark degrees
of freedom.  
Its other presence in the chiral lagrangian, imposed by $U_A(1)$,
involves couplings that are more suppressed,
$1/N_c^{\frac{3}{2}}$  per $\eta^0$ in
a vertex. This is associated to the
special gluonic content of the $\eta^0$, put forward by the anomaly.
Although there would be no such a thing as the $\eta^0$ in a world
without quarks - nor chiral symmetry -, in the chiral limit
the $\eta^0$ mass is \cite{w1}
$$\left. \left.  m_{\eta^0}^2\right|_{U_A(1)}
= \frac {4 n_l}{f^2} \left( \frac {d^2 E}{d \theta^2}
\right) \right|^{no \; quarks}_{\theta=0} + O(\frac {1}{{N_c}^2})$$ 
where $E$ is the vacuum energy
in a world without quarks and with a coupling
to the topological charge $Tr \ G^{\mu \nu} \tilde{G}_{\mu \nu}$
in the lagrangian, as in (\ref{2.6}). 
The fixed proportion of $\eta^0$ and $\hat{\theta}$ that appear in the
combination $X= i \frac {\sqrt{2 n_l}}{f} \eta^0 + \hat{\theta}$ actually
relates glueball and  $\eta^0$ anomalous couplings 
to operators involving goldstone bosons.
The vertex suppression of 
$1/N_c^{\frac{3}{2}}$ per $\eta^0$ is a combination of $1/N_c$-glueball
and $1/\sqrt{N_c}$-meson suppression, the former originated in the anomaly
equation, the second carried by the $f$ factor that divides the $\eta^0$ field.

\vspace*{0.5cm}

In order to obtain the $1/N_c$ power counting of the
sub-leading pieces too, one might proceed
by comparison of Green's functions, as evaluated from the
chiral lagrangian and from the diagrams in QCD. 
In the effective theory, given an operator 
multiplied by a function $G(X)$,
the $1/N_c$ power counting can be established on the basis of
two distinguishing features:
the number of traces over flavour indices ($\# (tf)$),
and the number of powers of the source $\hat{\theta}(x)$, ($\#(\hat{\theta}))$.
As a rule of thumb, the leading
dependence on $1/N_c$ of the various couplings follows from the
simple prescription \cite{l}: 
\begin{equation}
G(X)= N_c^{2-\#(tf)-\#(\hat{\theta})} \ g ( X/N_c ),
\label{4.1}
\end{equation}
where $g$ is a function whose expansion in powers of $X/N_c$ has coefficients
of order 1.
The origin of each factor can be easily traced:
in relation to the vacuum energy which is $O(N_c)^2$,
there is one power suppression of $1/N_c$ for each flavour trace and one for
each source - recall that in QCD the sources for $1/N_c$ suppression are
the loops of quarks and the non-planarity of the
diagrams.
Each trace taken over flavour indices amounts to a sum
over quark flavours, which in turn can arise only in a quark loop
in QCD and adds a factor of $1/N_c$. As for the
source $\hat{\theta}(x)$, it couples to the topological
charge in (\ref{2.6}) with strength $1/N_c$; each derivative
of the generating functional with respect to $\hat{\theta} (x)$ will bring
one power of $1/N_c$ to the Green's function. In the
effective theory this 
is achieved by pulling out an explicit power of $1/N_c$ for each 
$\hat{\theta}(x)$. Finally, it was already mentioned that
the ubiquitous factor of $f$ count as $\sqrt{N_c}$. In particular, 
the $\eta^0$ that appears in $X$ is always suppressed by a factor $1/f$ 
(\ref{3.5bis}).
There are no additional powers of $1/N_c$ for the leading contributions: once 
all the previous factors of $1/N_c$ have been pulled out,
also from $X$ as $X/N_c$, the expansion
of $g$ in powers of $X/N_c$ has coefficients that are order 1.

Recall at this point that the $1/N_c$ counting should be done in
a chiral lagrangian with a generic number $n_l$ of light flavours.
This is because of the $n_l$-dependent factorization relations already
mentioned in section 3, that give linear relations among the {\it a
priori} independent operators for particular values of $n_l$.
The mismatch of powers of
$1/N_c$ and the departure from the general rule (\ref{4.1}) are avoided
by allowing for a generic $n_l$. The correct counting is thus obtained
for the functions $\beta_i (X)$ in (\ref{3.22}),(\ref{3.19}). 
The implications for the $L_i(X)$ can be read off from (\ref{3.27}).
 
Furthermore, as pointed out in \cite{dRP}, it is 
the $U_L \otimes U_R$ lagrangian
that provides the suitable basis to establish the correct
$1/N_c$ power counting of the constants that involve the nonet of mesons. 
The $\eta'$ in the large-$N_c$ limit gives a contribution to the
large-distance behaviour of the Green's functions that should not be
overlooked; in the limit its properties are the same as for the rest of
goldstone bosons in the nonet. In next to leading order,
a topological mass term appears for the $\eta'$, but it is 
$O(1/N_c)$ and is treated perturbatively.
In counting powers of $1/N_c$, the $\eta'$ cannot be integrated out
from the nonet,
for it is when $N_c \to \infty$, when the $1/N_c$ expansion is more sensible,
that the $\eta'$ becomes massless and does not decouple.

Expanding the $W_i$'s functions in (\ref{3.7}) in power series in X,
\begin{equation}
W_k(X)=W_{k0} + W_{k2} X^2 + W_{k4} X^4 + ...= \frac {f^2}{4}
\left( v_{k0} + v_{k2} X^2 + v_{k4} X^4 + ...\right), 
\label{p}
\end{equation}
for $k=0,1,2,4,5,6$, and for $W_3$
\begin{equation}
W_3(X)=W_{31} X + W_{33} X^3 + \; ...= -i \frac {f^2}{4}
\left( v_{31} X + v_{33} X^3 + \; ...\right),
\label{d}
\end{equation}
and following rule (\ref{4.1}) we find
\begin{eqnarray}
W_{00}&=& O(N_c^2)        \nonumber\\
W_{10}&=& W_{20}\  =\ \frac {f^2}{4}\ =\ O(N_c)                \nonumber\\
W_{31}&=& O(1), \ \ \ \ W_{50}\ =\  O(1), \ \ \ \    W_{60}\  =\  O(1) 
.   \nonumber\\
\nonumber
\label{t}
\end{eqnarray}

For the coefficients of the $O(p^4)$ operators
that do not involve derivatives of $\hat{\theta}$ \cite{dRP},
for $n_l=3$ we find,

\begin{eqnarray}
L_1(0) \; ,L_2(0)\; ,L_3(0)\; ,L_5(0)\; ,L_8(0)\; ,L_9(0)\; ,L_{10}(0)\; ,
L_{11}(0) \; ,L_{12}(0) \; , \nonumber\\
L_{13}(0)\; ,L_{14}(0)\; ,L_{15}(0)\; ,L_{16}(0) &=& O(N_c) \nonumber\\
2L_1(0)-L_2(0)\; , L_{13}(0)+8 L_1(0)\; , L_{14}(0)-2 L_1(0)\; ,
L_4(0)\; , L_6(0)\; , \nonumber\\
L_7(0)\; ,L_{18}(0)\; , L_{19}(0)\; ,
L_{20}(0)  &=& O(1) \nonumber\\
L_{15}(0)-2L_{14}(0)\; , 2 L_{16}(0)- L_{14}(0)\; , L_{17}(0) &=& O(1/N_c).
\nonumber\\
\label{c}
\end{eqnarray}
For the parity-odd terms, the first contribution is given by the
linear term of the $X$ expansion, $L_i'(0)$:
\begin{eqnarray}
L'_{21}(0)\; , L'_{25}(0)\;  , L'_{28}(0)
&=& O(1) \nonumber\\
L'_{22}(0)\; ,L'_{23}(0)\; ,L'_{26}(0)\; ,L'_{27}(0) &=& O(1/N_c)
 \nonumber\\
L'_{24}(0) &=& O(1/N_c^2).
\nonumber\\
\label{cp}
\end{eqnarray}

As mentioned, the $\eta^0$ gets a contribution to its mass
that is $O(1/N_c)$, in the notation of (\ref{d}) $\left. m^2_{\eta^0}
\right|_{U_A(1)}= - {n_l} v_{02}$. Counting $m^2_{\eta^0}$
as two chiral powers $O(p^2)$, in the multiple expansion we shall 
count $1/N_c$ also as $O(p^2)$ \cite{l}.
However, to be fully consistent with it, if terms $O(p^4) \times 
\frac {1}{N_c}$ are kept, then the chiral $O(p^6)$ order should
be also included. This would require a two-loop chiral perturbation theory
calculation which is far beyond the scope of this article.

Following these criteria and using (\ref{p}), (\ref{d}) and (\ref{t}), 
we expand the $B_i$ and $\Omega_i$ functions from Appendix D, first 
in powers of $X$ and then in 
powers of $\frac{1}{N_c}$, keeping corrections  up to $\frac{1}{N_c}$ for 
the $O(p^4)$ terms, up to $\frac{1}{{N_c}^2}$ for the $O(p^2)$ terms and 
up to $\frac{1}{{N_c}^3}$ for the $O(p^0)$ one. Notice that none of the
$\eta^0$ fields that appeared through the $X$ has survived, which means that
all the terms in the lagrangian starting with more that four fields are 
eliminated.
This kind of terms would be only required for calculating processes that are
very difficult to measure experimentally.

To this order, only two $\Omega_i$'s survive:
\begin{equation}
\begin{array}{ll}
\Omega_0 = \frac {\nz^2}{2} \vzd ^2 + O(\frac{1}{{N_c}^4})  , \;\;\;&
\Omega_2 = -  \frac {1}{2}\vzd + \nz\vzd\vtu + O(\frac{1}{{N_c}^3}).\nonumber\\
\nonumber
\end{array}
\nonumber
\end{equation}
The $B_i$'s for i=0 to 20 are the same as in (\ref{llista}) except for
two new contributions:
\begin{equation}
\begin{array}{ll}
B_8 = \frac {\nz}{16}-  \frac {1}{2}\vtu + O(\frac{1}{{N_c}^2}), \;\;\;&
B_{12} =  \frac {\nz}{8} -\vtu + O(\frac{1}{{N_c}^2}),  \nonumber \\ 
\nonumber
\end{array}
\nonumber
\end{equation}

The rest of coefficients either vanish exactly or do not contribute 
to this order in the expansion. There are also some contributions
that are proportional to $\hat{\theta}$.
\vspace*{0.5cm}

The use of the equations of motion does not ruin the $1/N_c$ power counting
in (\ref{4.1}), although it introduces additional products 
of traces and powers of derivatives of $\hat{\theta}$ in (\ref{3.10}).
The very structure of (\ref{3.10}) complies
with (\ref{4.1}) since the functions of $X$ that multiply the various
operators carry their own
$1/N_c$ suppression required by
(\ref{4.1}): for the identity,
$\frac {W_0'}{W_1}= g(X/N_c)$; $\langle C^\mu C_\mu\rangle$,
$\langle C^\mu \rangle \ C_\mu$ and $D^\mu \hat{\theta} \ C_\mu$ are multiplied
by $\frac {W_1'}{W_1} = \frac {1}{N_c} \ g(X/N_c)$; $N$ and $M$ do
not get any suppression since both $\frac {W_2}{W_1}$ and $\frac {W_3}{W_1}$
are $g(X/N_c)$; however, $\langle M \rangle$ and $\langle N \rangle$
appear multiplied by $\frac {W_2'}{W_1}$ and $\frac {W_3'}{W_1}$ which are
$\frac {1}{N_c} \ g (X/N_c)$; $\frac {W_2}{W_1}=\frac {1}{N_c} \ 
g (X/N_c)$ multiplies $D^\mu D_\mu \theta $ whereas 
$\frac {W_5'- W_6'}{W_1}=\frac {1}{{N_c}^2} \ g(X/N_c)$. Here $g$
denotes, in each case, some function that has $N_c$-independent Taylor
expansion coefficients.

Finally, let us comment on the regeneration through quantum corrections
of a term \break
$\langle U^\dagger D_\mu U\rangle 
\langle U^\dagger D^\mu U\rangle$, which had been 
removed from the tree level effective lagrangian in (\ref{3.7}).
It is a confirmation that the effects of meson loops
bring to the effective action contributions that are $1/N_c$ suppressed 
in relation to the leading tree level. 
It is only at $N_c =\infty$ that $\pi$, $K$, $\eta$,
$\eta'$ form a nonet. When sub-leading corrections in $1/N_c$ are taken into
account, the enlarged symmetry with respect to $U_L \otimes U_R$ no longer 
holds.
The first instance is the $O(1/N_c)$ mass piece
exclusive for the $\eta_0$. The reappearance of that term is another example:
the $\eta^0$ (singlet) and the $\pi$ (octet)
fields are normalized differently by sub-leading contributions.

\section{Conclusion and outlook.}

In this article the effective theory developed in \cite{gl} for the 
strong interactions at low energies among the lightest pseudoscalars
is extended
to include the $\eta'$ particle. The approach that has been adopted
exploits the fact that, as the number of colours $N_c$ grows bigger,
the mass difference (topological mass) between the $\eta '$ and the
octet vanishes as $1/N_c$, and the $U_A(1)$ puzzle can be treated as
a series in inverse powers on $N_c$. We exploit the possibility
that a good description could emerge by taking the
nonet of soft pseudoscalars as the goldstone bosons of the spontaneous 
breaking of $U_L \otimes U_R \to U_V$, 
thus benefitting from the perks that such theories feature, 
in terms of constraints on the form of the couplings and
relations among the coupling constants. The departures from the results
in the real world are dealt with as corrections
in powers of the quark masses and $1/N_c$.

The most general effective action is given that includes
terms up to $O(p^4)$, with quantum corrections included to one-loop.
The $U_A(1)$ anomaly is conveniently incoporated.

This is the first step of a project towards a systematic study to spell out
the predictions that could emerge from such scenario. It will shed light
on what processes can be accomodated within the approach.

%
%
%

\section{Acknowledgments}

We are grateful to D. Espriu,
E. de Rafael and S. Peris for discussions as well as
R. Tarrach for his critical reading of the manuscript.
Financial support from CICYT, contract AEN95-0590,
and from CIRIT, contract GRQ93-1047 are
acknowledged. J.T. acknowledges the Benasque Center for Physics
and the Physics Department at Brookhaven National Laboratory for the
hospitality extended to him during the completion on this
work. P.H.-S. acknowledges a Grant from the {\it Generalitat de Catalunya}.

\newpage

\section{Appendix A: The Heat-Kernel Technique.}

    This appendix is a reminder of some results that have
been used in the text, which  are obtained with the help
the heat-kernel  technique. It provides a convenient way to evaluate
the one-loop effective action.

Let $\hat{H}$ be an operator acting on a Hilbert space.
The heat-kernel two-point function associated to $\hat{H}$
is defined as a function of the parameter $\tau$, 
$$K(x,y;\tau)=\langle x|e^{-i\tau \hat{H} }|y\rangle  \theta(\tau).$$
It verifies the equation
$$i\frac {\partial}{\partial \tau} K(x,y;\tau)= \int d^D z \;\langle x|\hat{H}|z\rangle 
K(z,y;\tau) +i \delta (\tau) \delta^D (x-y),$$
with the boundary condition,
$$ \lim_{\tau \to 0} K(x,y;\tau) = \delta^{D}(x-y),$$
which follows from its definition. We do not specify the dimensionality
of space-time and allow for a generic $D$.
Often, $\hat{H}$ is a {\it local} operator, i.e.,
$\langle x|\hat{O}|y\rangle =O_x \delta^{D}(x-y)$, where $O_x$ is a 
differential
operator with a finite number of terms. We shall limit ourselves to this case.
Then, $K(x,y; \tau)$ is a Green's function that verifies
a  partial differential equation of the Schr\"odinger type,
$$i\frac {\partial}{\partial \tau} K(x,y;\tau)= O_x K(x,y;\tau)+i \delta (\tau) \delta^D (x-y).$$

When the operator $\hat{H}$ is {\it elliptic} the definition
of $K$ differs somewhat from the one given above. In that case there
is no need for a factor of $i$ in the exponent, since all the eigenvalues
of $\hat{H}$ are non-negative. There, the equation is a
heat-transport-like equation, from which the technique shares its
name. In the present case we are dealing with 
operators of the sort $\partial_\mu \partial^\mu$, which in Minkowski
space is {\it hyperbolic}.
It is convenient to pick out the mass term from $\hat{H}$, if the theory
is massive, or, else, to introduce an infrared regulator by adding a
constant term $M^2$ to $\hat{H}$, $\hat{H}+ M^2$.
For small values of $\tau$, $K(x,y;\tau)$ admits an asymptotic expansion
of the form,
\begin{equation}
K(x,y;\tau) \sim i \frac {1}{(4 \pi i \tau)^{\frac {D}{2}}} \ \
e^{-i M^2 \tau + \frac {(x-y)^2}{4 i \tau}}\ \sum_{n=0}^{\infty} h_n(x,y)
(i\tau)^n.
\label{SDW}
\end{equation}
The functions $h_n(x,y)$ are known as the {\it Seeley-DeWitt} coefficients
\cite{ball}.

Given that
$$K(x,y;\tau \to 0) \sim i \frac {1}{(4 \pi i  \tau)^{\frac {D}{2}}}\
e^{\frac {(x-y)^2}{4 i \tau}}\ \longrightarrow\ \delta^{D}(x-y),$$
the boundary condition translates into
$$h_0(x,x)=1.$$

The computation of the one-loop effective action using the background field
method entails the evaluation of
$$\log \det (\hat{H})= Tr \; \log \hat{H}.$$
$Tr$ stands for the trace over space-time as well as all the internal
indices. This can be written in terms of the kernel $K(x,y; \tau)$ if one uses
the following integral representation for the logarithm
$$\log \hat{H} -\log \hat{H}_0= - \int_0^{\infty} \frac {d \tau}{\tau}
\left( e^{-i \tau \hat{H} }
-e^{-i \tau \hat{H}_0 } \right).$$
Apart from an unessential $\hat{H}$-independent, divergent constant $C$
\begin{equation}
Tr\; \log \hat{H}= -\int d^Dx \int_0^{\infty} \frac {d \tau}{\tau}
\ tr \ \langle x|e^{-i \tau \hat{H} }|x\rangle  + \ C.
\label{heatk}
\end{equation}
where tr stands for the trace over internal indices only.

This expansion will be used to depict the divergent short-distance
contributions so as to be able to subtract them away.
Given $\tau$, only those points separated an interval
$(x-y)^2$ that is less or of the order of
$\tau$ give non-negligible (non-oscillatory)
values for $K(x,y;\tau)$. The
short distance contribution to the effective action is thus contained
in the integral about $\tau=0$; it will not be 
affected by $M$, which only modifies the contribution from large
values of $\tau$ (large wavelength modes).

The particular form of the operator $\hat{H}$ which corresponds to 
our computation is 
$$\hat{H}= d_\mu d^\mu +\sigma\ ,$$
where $d_\mu= \partial_\mu +\omega_\mu (x)$.
A na\"{\i}ve application of eq. (\ref{heatk}) could lead to infrared
divergences associated to the large $\tau$ integration region.
For that reason we have changed it to $\hat{H}+M^2= d_\mu d^\mu +\sigma+M^2$.
Each term in the Seeley-DeWitt expansion yields a finite contribution,
which give 
\begin{equation}
i \ tr \langle x|\log \left( \hat{H} + M^2 \right)|x\rangle =  
\frac {1}{(4 \pi)^{D/2}} \sum_{n=0}^{\infty}
(M^2)^{\frac {D}{2} -n} \Gamma\left(n - \frac {D}{2} \right) 
\langle h_n(x,x) \rangle,
\label{a.1}
\end{equation}
except for an unessential additive constant. The Seeley-DeWitt coefficients 
\footnote{In general $\sigma (x)$ and 
$\omega (x)$ can be matrices. In that case
the coefficients $h_n$ are matrices as well and a trace over these internal
indices is also understood in (\ref{a.1}).}.
are known in this case,
\begin{eqnarray}
h_0(x,x) &=& I, \nonumber \\
h_1(x,x) &=&- \sigma, \nonumber \\
h_2(x,x) &=& \frac {1}{2} \sigma^2 + \frac {1}{12} R_{\mu \nu} R^{\mu \nu}
+\frac {1}{6} d_\mu d^\mu \sigma.
\label{a.3}
\end{eqnarray}
where 
$R_{\mu \nu}=\partial_\mu \omega_\nu - \partial_\nu \omega_\mu+[\omega_\mu,
\omega_\nu].$
Higher order $h_n(x,x)$ may be found in \cite{ball}.

The singularities that arise in the calculation have been regulated
by analytic continuation over the $D$-complex plane.
In a four-dimensional
theory the ultraviolet divergences appear as poles about $D=4$, 
and get contributions from $h_0$, $h_1$ and $h_2$,
as $\Gamma  \left( - \frac {D}{2} \right)$, 
$\Gamma  \left( 1 - \frac {D}{2} \right)$ and 
$\Gamma  \left( 2- \frac {D}{2} \right)$, respectively. Retaining in 
(\ref{a.1})
these three terms only, it reads
\begin{eqnarray}
\left.
i \ tr \langle x|\log \left( \hat{H} + M^2 \right)|x\rangle 
\right|_{div} &=&
\frac {M^{D-4}}{(4 \pi)^2} \left(  \frac {-2}{D-4} +
\log 4 \pi - \gamma +1 \right)
\langle  \ \frac {1}{2} (\sigma + M^2)^2 +
\frac {1}{12} R_{\mu \nu} R^{\mu \nu}  \ \rangle 
\nonumber \\
&+& \frac {1}{(4 \pi)^2} \langle \ \left( \frac {1}{4} M^4
- \frac {1}{2} \sigma^2 \right) - 
\frac {1}{12}  R_{\mu \nu} R^{\mu \nu} \ \rangle + O(D-4), 
\label{a.4}
\end{eqnarray}

Notice that in (\ref{a.4}) it is the combination 
$\sigma + M^2 $ the one that multiplies the pole $\frac {1}{D-4}$, i.e., it
is the whole operator that adds to 
$d_\mu d^\mu$ in $\hat{H}+M^2$.
If $\sigma$ has its own mass terms - the light pseudoscalar masses 
in our case -
there is no need to introduce any new infrared regulator $M$. 
We see that only $h_2$ is involved in the residue of the pole
$\frac {1}{D-4}$.

Therefore, a finite expression is obtained by subtracting
$$\frac {\mu^{D-4}}{(4 \pi)^2} \left( - \frac {2}{D-4} +
\log 4 \pi - \gamma +1 \right)
\langle  \ h_2(x,x) \ \rangle$$
to the effective action. This copes with the ultraviolet divergences
and it is the procedure used in the text.
$\mu$ is a parameter with {\it mass} units. 
The total derivative 
$\langle d_\mu d^\mu \sigma \rangle$ has been discarded.

%


\section{Appendix B: U(3).}

This appendix is devoted to present some of the properties of the $U(3)$
group which have been used in the text. Whenever the generalization is 
straightforward we give the result for $U(n)$. 

The explicit form of the $U(3)$ hermitian generators
, $\lambda_\mu=\lambda_\mu^\dagger$
, $\mu=0,1,2,...,8$, is 
\begin{eqnarray}
\label{unmatrices}
\lambda_0 =\sqrt{\frac {2}{3}}
\left(\begin{array}{rrr} 
1 & 0 & 0 \\
0 & 1 & 0 \\
0 & 0 & 1 \end{array}\right),\
&
\lambda_1 =  \left(\begin{array}{rrr}
0 & 1 & 0 \\
1 & 0 & 0 \\
0 & 0 & 0 \end{array}\right),
&
\lambda_2 =  \left(\begin{array}{rrr}
0 & -i & 0 \\
i & 0 & 0 \\
0 & 0 & 0 \end{array}\right),
\\
     &    &  \nonumber   \\
     &    &  \nonumber   \\
\lambda_3 = \left(\begin{array}{rrr}
1 & 0 & 0 \\
0 & -1 & 0 \\
0 & 0 & 0  \end{array}\right),\nonumber
&
\lambda_4 =  \left(\begin{array}{rrr}
0 & 0 & 1 \\
0 & 0 & 0 \\
1 & 0 & 0  \end{array}\right),
&
\lambda_5 =  \left(\begin{array}{rrr}
0 & 0 & -i \\
0 & 0 & 0 \\
i & 0 & 0  \end{array}\right),
\\
     &   &  \nonumber    \\
     &   &  \nonumber    \\
\lambda_6 = \left(\begin{array}{rrr}
0 & 0 & 0 \\
0 & 0 & 1 \\
0 & 1 & 0 \end{array}\right), \nonumber
&
\lambda_7 =  \left(\begin{array}{rrr}
0 & 0 & 0 \\
0 & 0 & -i \\
0 & i & 0  \end{array}\right),
&
\lambda_8 =  \frac {1}{\sqrt{3}}\left(\begin{array}{rrr}
1 & 0 & 0 \\
0 & 1 & 0 \\
0 & 0 & -2 \end{array}\right).
\end{eqnarray}
For $U(n)$ there are $n^2$ such matrices, and $\mu= 0,1,...,n^2-1$.
Also $\lambda_0=\sqrt{\frac {2}{n}} I$, with $I$ the $n \times n$ identity.
These matrices obey, in general, the following basic trace properties:
\begin{equation}
\label{basic}
Tr\left(\lambda_\mu\right)\equiv \langle  \lambda_\mu\rangle =
\sqrt{2n}\ \delta_{\mu 0}
\qquad,\qquad
\langle \lambda_\mu\lambda_\nu\rangle =2 \delta_{\mu\nu}\ . 
\end{equation}
The product of two matrices verifies:
\begin{equation}
\left[\lambda_\mu,\lambda_\nu\right]= 2 i f_{\mu\nu\rho}\lambda_\rho\qquad
,\qquad
\left\{\lambda_\mu,\lambda_\nu\right\}= 2 d_{\mu\nu\rho}\lambda_\rho\ ,
\end{equation}
\begin{equation}
\lambda_\mu\lambda_\nu=\left( d_{\mu\nu\rho}+ i f_{\mu\nu\rho}\right) 
  \lambda_\rho\ .
\end{equation}
For $U(3)$, the non-zero entries of the antisymmetric
$f_{\mu\nu\rho}$ and symmetric $d_{\mu\nu\rho}$ constants are
\begin{eqnarray}
&f_{123}=1\qquad, \qquad f_{458}=f_{678}={\sqrt{3}\over 2} \ ,\nonumber\\
&f_{147}=-f_{156}=f_{246}=f_{257}=f_{345}=-f_{367}=\frac {1}{2} \ ,
\end{eqnarray}
\begin{eqnarray}
&d_{0\mu\nu}=\sqrt{2\over 3} \delta_{\mu\nu}\ ,\nonumber\\
&d_{118}=d_{228}=d_{338}=-d_{888}={1\over \sqrt{3}}\ ,\nonumber\\
&d_{146}=d_{157}=-d_{247}=d_{256}=d_{344}=d_{355}=-d_{366}
=-d_{377}={1\over 2}\ ,\nonumber\\
&d_{448}=d_{558}=d_{668}=d_{778}=-{1\over 2\sqrt{3}}\ .
\end{eqnarray}
These tensors are not independent as they satisfy the following relations
(repeated indices are summed up):
\begin{eqnarray}
& d_{\mu\nu\nu}=n \sqrt{2n}\ \delta_{\mu 0}\ , \nonumber\\
\nonumber\\
& d_{\mu\nu\lambda}d_{\rho\nu\lambda}= 
n\left(\delta_{\mu\rho}+\delta_{\mu 0}
\delta_{\rho 0}\right)\ ,\nonumber\\
& f_{\mu\nu\lambda}f_{\rho\nu\lambda}= 
n\left(\delta_{\mu\rho}-\delta_{\mu 0}
\delta_{\rho 0}\right)\ , \nonumber\\
\nonumber\\
&f_{\mu\nu\tau} f_{\lambda\rho\tau}+f_{\mu\lambda\tau}f_{\rho\nu\tau}+
f_{\mu \rho\tau}f_{\nu\lambda\tau}=0\ ,\nonumber \\
&f_{\mu\nu\tau} d_{\lambda\rho\tau}+f_{\mu\lambda\tau}d_{\rho\nu\tau}+
f_{\mu \rho\tau}d_{\nu\lambda\tau}=0\ ,\nonumber \\
\nonumber\\
&f_{\mu\nu\sigma}f_{\rho\tau\sigma}=d_{\mu\rho\sigma}d_{\tau\nu\sigma}-
d_{\mu\tau\sigma}d_{\nu\rho\sigma}\ .
\end{eqnarray}
The identity
\begin{equation}
\left(\lambda_\alpha\right)_{a b}
\left(\lambda_\alpha\right)_{c d}= 2 \delta_{a d}\delta_{b c}\ ,
\end{equation}
has been extensively used. It yields the properties
\begin{eqnarray}
\lambda^{\alpha} \lambda^\alpha &=& 2 n I , \nonumber \\
\langle  \lambda_\alpha A \lambda_\alpha B\rangle & =& 2 \langle  A\rangle 
\langle  B\rangle \ ,\nonumber \\
\langle  \lambda_\alpha A\rangle  \langle 
\lambda_\alpha B\rangle  &=& 2 \langle  A B\rangle  \ .
\end{eqnarray}

\section{Appendix C: Operators with derivatives of $\hat{\theta}$ and 
some \\ C-violating operators.}

In this appendix to give list of the operators that involve derivatives 
of $\hat{\theta}$, that we have not included in the text 
(see (\ref{3.18})).

\begin{eqnarray}
O_{31} &=& D_\mu \hat{\theta}\ \langle C^\mu C^\nu C_\nu\rangle , \nonumber\\
O_{32} &=& D_\mu \hat{\theta}\ \langle C^\mu\rangle \langle C^\nu C_\nu\rangle , \nonumber\\
O_{33} &=& D_\mu \hat{\theta}\ \langle C^\mu C^\nu\rangle \langle C_\nu\rangle , \nonumber\\
O_{34} &=& D_\mu \hat{\theta}\ \langle C^\mu\rangle \langle C^\nu\rangle \langle C_\nu\rangle , \nonumber\\
O_{35} &=& D_\mu \hat{\theta}\ D^\mu \hat{\theta}\ \langle C^\nu C_\nu\rangle , \nonumber\\
O_{36} &=& D_\mu \hat{\theta}\ D_\nu \hat{\theta}\ \langle C^\mu C^\nu\rangle , \nonumber\\
O_{37} &=& D_\mu \hat{\theta}\ D^\mu \hat{\theta}\ \langle C^\nu\rangle \langle C_\nu\rangle , \nonumber\\
O_{38} &=& D_\mu \hat{\theta}\ D_\nu \hat{\theta}\ \langle C^\mu\rangle \langle C^\nu\rangle , \nonumber\\
O_{39} &=& D_\mu \hat{\theta}\ D^\mu \hat{\theta}\ D_\nu \hat{\theta}
\langle C^\nu\rangle , \nonumber\\
O_{40} &=& D_\mu \hat{\theta}\ D^\mu \hat{\theta}\ D_\nu \hat{\theta}\
D^\nu \hat{\theta} ,\nonumber\\
O_{41} &=& i D^\mu D_\mu \hat{\theta}\ \langle C^\nu C_\nu\rangle  ,\nonumber\\
O_{42} &=& i D^\mu D_\mu \hat{\theta}\ \langle C^\nu\rangle \langle C_\nu\rangle  ,\nonumber\\
O_{43} &=& i \langle C^\mu\rangle D_\mu \hat{\theta}\ D^\nu D_\nu \hat{\theta},\nonumber\\
O_{44} &=& i D_\mu \hat{\theta}\ D^\mu \hat{\theta}\ D^\nu D_\nu \hat{\theta}
,\nonumber\\
O_{45} &=& D^\mu D_\mu \hat{\theta}\ D^\nu D_\nu \hat{\theta} ,\nonumber\\
O_{46} &=& D_\mu \hat{\theta}\ \langle C^\mu M\rangle  ,\nonumber\\
O_{47} &=& D_\mu \hat{\theta}\ \langle C^\mu\rangle \langle M\rangle  ,\nonumber\\
O_{48} &=& i D_\mu \hat{\theta}\ \langle C^\mu N\rangle  ,\nonumber\\
O_{49} &=& i D_\mu \hat{\theta}\ \langle C^\mu\rangle \langle N\rangle  ,\nonumber\\
O_{50} &=& D_\mu \hat{\theta}\ D^\mu \hat{\theta}\ \langle M\rangle  ,\nonumber\\
O_{51} &=& i D_\mu \hat{\theta}\ D^\mu \hat{\theta}\ \langle N\rangle  ,\nonumber\\
O_{52} &=& i D^\mu D_\mu \hat{\theta}\ \langle M\rangle  ,\nonumber\\
O_{53} &=& D^\mu D_\mu \hat{\theta}\ \langle N\rangle  ,\nonumber\\
O_{54} &=& D_\mu \hat{\theta}\ \langle C_\nu \left( F_L^{\mu \nu} - 
U^\dagger F_R^{\mu \nu} U \right)\rangle  ,\nonumber\\
O_{55} &=& D_\mu \hat{\theta}\ \langle C_\nu\rangle \langle  F_L^{\mu \nu} - F_R^{\mu \nu}\rangle ,
\nonumber\\
O_{56} &=& \epsilon_{\mu \nu \rho \sigma}
\langle  \left( F_L^{\mu \nu} - 
U^\dagger F_R^{\mu \nu} U \right) C^\rho\rangle  D^\sigma \hat{\theta}
,\nonumber \\
O_{57} &=& \epsilon_{\mu \nu \rho \sigma}
\langle  F_L^{\mu \nu} - F_R^{\mu \nu}\rangle  \langle C^\rho\rangle  \ 
D^\sigma \hat{\theta}.
\label{Dtheta}
\end{eqnarray}

The left-over independent operators are $C$-violating and are the following:

\begin{equation}
\begin{array}{ll}
\langle  C_\mu\rangle 
\langle \left( F_L^{\mu \nu} + U^\dagger F_R^{\mu \nu} U \right) 
C_\nu\rangle , 
& i \langle C_\mu C_\nu \left( F_L^{\mu \nu} - U^\dagger F_R^{\mu \nu} U
\right)\rangle  ,
\nonumber \\ \langle F_L^{\mu \nu}F_{L \;\mu \nu}-F_R^{\mu \nu}
F_{R \;\mu \nu}\rangle ,
& \langle F_L^{\mu \nu}\rangle \langle F_{L \;\mu \nu}\rangle -
\langle F_R^{\mu \nu}\rangle \langle F_{R \;\mu \nu}\rangle ,
\nonumber \\ \epsilon_{\mu \nu \rho \sigma} \langle C^\mu\rangle \langle 
C^\nu C^\rho C^\sigma\rangle ,
& \epsilon_{\mu \nu \rho \sigma} \langle 
 \left( F_L^{\mu \nu} + 
U^\dagger F_R^{\mu \nu} U \right) C^\rho\rangle \langle C^\sigma\rangle  ,
\nonumber\\
D_\mu \hat{\theta}\ \langle C_\nu \left( F_L^{\mu \nu} + 
U^\dagger F_R^{\mu \nu} U
\right)\rangle  ,
& D_\mu \hat{\theta}\ \langle C_\nu\rangle \langle  F_L^{\mu \nu} + 
F_R^{\mu \nu}\rangle , 
\nonumber\\ i \epsilon_{\mu \nu \rho \sigma}\ D^\mu \hat{\theta}\ 
\langle C^\nu C^\rho C^\sigma\rangle  ,
& \epsilon_{\mu \nu \rho \sigma} 
\langle  \left( F_L^{\mu \nu} + 
U^\dagger F_R^{\mu \nu} U \right) C^\rho\rangle  D^\sigma \hat{\theta},
\nonumber\\ 
\epsilon_{\mu \nu \rho \sigma} 
\langle  F_L^{\mu \nu} + F_R^{\mu \nu}\rangle \langle C^\rho\rangle  \
D^\sigma \hat{\theta} &
\end{array}
\nonumber
\end{equation}

\section{Appendix D:  The one-loop renormalization functions} 

In this appendix, we collect the functions $\Omega_i$'s and $B_i$'s, as
defined in equation (\ref{3.19}) , which have been computed with the heat-
kernel method. The results have been obtained with the help of the
program {\sl Mathematica}.

The $O(p^0)$ lagrangian is renormalized with $\Omega_0$, 
whereas the rest of the 
$\Omega_i$'s renormalize the $O(p^2)$. For technical reasons, it was
convenient to keep one more operator at $O(p^2)$ than those given in 
(\ref{3.22p}), $E_7 = iD_\mu D^\mu \hat{\theta}$. By integration by parts,
this can be written in terms of the other $E_i$, 
$$\Omega_7 E_7 = - i\Omega_7' E_5 - i\Omega_7' E_6,$$
which should be taken into account and added to $\Omega_5$ and $\Omega_6$.

For the sake of brevity, we use the following short notation: 
$$\omega_i \equiv \frac {W_i}{W_1}, \\\\\
\omega_i^{\prime} \equiv \frac {W_i ^{\prime }}{W_1}, \\\\\
\omega_i^{\prime \prime} \equiv \frac {W_i ^{\prime \prime}}{W_1},$$
and, similarly, to include as many
primes as necessary. Let us emphasize that $\omega_i^{\prime}$ {\it is
not} the derivative of $\omega_i$, and so forth.
Also, recall that $W_1(0)=W_2(0)=\frac {f^2}{4}$ so that $\omega_1(0)=
\omega_2 (0)=1$.
\begin{eqnarray}
\Omega_0 &=& \frac {\nz^2}{8} \wopp^2 - \frac {\nz^2}{8} \wop \wopp \wunp
+ \frac {\nz^4}{32} \wop^2 \wunp^2   \nonumber \\
\Omega_1 &=& - \frac {\nz^2}{8} \wop \wunp - \frac {\nz^2}{4} \wopp \wunpp
+ \frac {\nz^2}{4} \wopp \wunp^2 + \frac {\nz^2}{8} \wop \wunp \wunpp
- \frac{\nz^4}{16} \wop \wunp^3  , \nonumber \\
\Omega_2 &=& - \frac {1}{4} \wopp \wdos - \frac {\nz^2}{4} \wopp \wdospp 
+ i \frac {\nz}{2} \wopp \wtresp + \frac {1}{4} \wop \wunp \wdos  
- \frac {\nz^2}{8} \wop \wunp \wdos  \nonumber \\ &+& 
\frac {\nz^2}{8} \wop \wunp \wdospp  +   
\frac {\nz^2}{8} \wopp \wunp \wdosp - i\frac {\nz}{4} \wop \wunp \wtresp - 
i\frac {\nz}{8} \wopp \wunp \wtres  \nonumber \\ &-& 
\frac {\nz^4}{16} \wop \wunp^2 \wdosp +
i\frac {\nz^3}{16} \wop \wunp^2 \wtres , \nonumber  \\
\Omega_3 &=& - i \frac {\nz}{2} \wopp \wdosp - \frac {1}{4} \wopp \wtres - 
 \frac {\nz^2}{4}\wopp \wtrespp + i \frac {\nz}{8} \wopp \wunp \wdos +
 i \frac {\nz}{4}\wop \wunp \wdosp  \nonumber \\ &+& 
\left( \frac {1}{4} - \frac {\nz^2}{8} \right) \wop \wunp \wtres + 
\frac {\nz^2}{8} \wop \wunp \wtrespp + \frac {\nz^2}{8} \wopp \wunp \wtresp -
\frac {\nz^4}{16} \wop \wunp^2 \wtresp  \nonumber \\ &-&  i \frac {\nz^3}{16}
\wop \wunp^2 \wdos , \nonumber \\ 
\Omega_4 &=& - \frac {\nz}{8} \wop \wunp  - \frac {\nz}{4} \wopp \wunpp +
\frac {\nz}{4} \wopp \wunp^2  +
\frac {\nz^3}{8} \wop \wunp \wunpp +
\left( \frac {\nz}{8}- 3 \frac {\nz^3}{16} \right)
\wop \wunp^3 , \nonumber \\ 
\Omega_5 &=& -\frac {1}{4} \left( \nz^3 - \nz \right) \wop \wunp^3 +
\frac {1}{4} \left( \nz^3 - \nz \right) \wop \wunp \wunpp , \nonumber \\
\Omega_6 &=& \frac {\nz}{4} \wopp \wunpp +  \frac {\nz^2}{4} \wopp \wcincpp -
\frac {\nz^2}{4} \wopp \wsispp  - \frac {\nz}{8} \wopp \wunp^2 +
\left( \frac {\nz^3}{8} - \frac {\nz}{4} \right)
 \wop \wunp \wunpp  \nonumber \\ &-&  \frac {\nz^2}{8} \wopp \wunp \wcincp -
  \frac {\nz^2}{8} \wop \wunp \wcincpp + 
\frac {\nz^2}{8} \wopp \wunp \wsisp +  \frac {\nz^2}{8} \wop \wunp \wsispp 
 \nonumber \\ &+&
\left( \frac {\nz}{8} - \frac {\nz^3}{16} \right) \wop \wunp^3 + 
  \frac {\nz^4}{16} \wop \wunp^2 \wcincp - 
\frac {\nz^4}{16} \wop \wunp^2 \wsisp , \nonumber \\ 
\Omega_7 &=& -i\frac {\nz}{4} \wopp \wunp -i  \frac {\nz^2}{4} \wopp \wcincp -i
\left( \frac {\nz^3}{8} - \frac {\nz}{4} \right) \wop \wunp^2 +i
\frac {\nz^2}{8} \wopp \wunp \wcinc \nonumber \\ &+&i 
\frac {\nz^2}{8} \wop \wunp \wcincp -i
\frac {\nz^4}{16} \wop \wunp^2 \wcinc .
\end{eqnarray}

\vspace*{0.5cm}
\noindent
The $O(p^4)$ is renormalized with the following $B_i$'s,
\begin{eqnarray}
B_0 &=& \frac {\nz}{48} + \frac {\nz}{6} \wunp^2 ,
\nonumber \\
B_1 &=& \frac {1}{16} + \frac {\nz^2}{8} \wunp^2 +  \frac {\nz^2}{8} \wunpp^2 -
\frac {\nz^2}{4} \wunp^2 \wunpp + \left( \frac {\nz^2}{48} + \frac {\nz^4}{32}
 \right) \wunp^4, \nonumber \\
B_2 &=& \frac {1}{8} + \frac {\nz^2}{24} \wunp^4 , \nonumber \\
B_3 &=&  \frac {\nz}{24} - \frac {\nz}{6} \wunp^2  , \nonumber \\
B_4 &=& \frac {\wdos}{8} - i \frac {\nz}{8} \wunp \wtres +
\frac {\nz^2}{4} \wunpp \wdospp + \frac {1}{4} \wunpp \wdos + 
\frac {\nz^2}{8}  \wunp \wdosp - i \frac {\nz^3}{2} \wunpp \wtresp  
\nonumber\\
&+& \left( \frac {\nz^2}{8} - \frac {3}{8}  \right) \wunp^2 \wdos -
\frac {\nz^2}{8} \wunp \wunpp \wdosp - \frac {\nz^2}{4} \wunp^2 \wdospp +
i \frac {\nz}{8} \wunp \wunpp \wtres  \nonumber\\
&+& i \frac {\nz}{2} \wunp^2 \wtresp +
\frac {\nz^4}{16} \wunp^3 \wdosp -
i \frac {\nz}{16} \wunp^3 \wtres , \nonumber \\
B_5 &=&  \frac {\nz}{8} \wdos - \frac {\nz}{8} \wunp^2 \wdos , \nonumber \\
B_6 &=& \frac {1}{16} \wdos^2 + \frac {1}{4} \wdos \wdospp + 
\frac {\nz^2}{8} \wdospp^2 - \frac {1}{4} \wtresp^2 - 
i \frac {\nz}{2} \wdospp \wtresp +
\left( \frac {\nz^2}{8} - \frac {1}{4} \right) \wunp \wdos \wdosp 
 \nonumber \\ &-& 
\frac {\nz^2}{8} \wunp \wdosp \wdospp - i \frac {\nz}{8} \wunp \wdos \wtres +
i \frac {\nz}{8} \wunp \wdospp \wtres + i \frac {\nz}{4} \wunp \wdosp \wtresp 
\nonumber \\ &+& \frac {\nz^4}{32} \wunp^2 \wdosp^2 + 
\left( \frac {1}{16}- \frac {\nz^2}{32} \right) \wunp^2 \wtres^2 -
i \frac {\nz^3}{16} \wunp^2 \wdosp \wtres , \nonumber \\
B_7 &=& \frac {1}{4} \wdosp^2 - \frac {1}{16} \wtresp^2 - 
\frac {\nz^2}{8} \wtrespp^2 - \frac {1}{4} \wtres \wtrespp -  
i \frac {\nz}{2} \wdosp \wtrespp - i \frac {\nz}{8}  \wunp \wdos \wtres 
 \nonumber \\ &+& i \frac {\nz}{4}  \wunp \wdosp \wtresp +
\left( \frac {1}{4}- \frac {\nz^2}{8} \right) \wunp \wtres \wtresp +
i \frac {\nz}{8} \wunp \wdos \wtrespp + \frac {\nz^2}{8} \wunp \wtresp \wtrespp
 \nonumber \\ &+& 
\left( \frac {\nz^2}{32} - \frac {1}{16} \right) \wunp^2 \wdos^2 -
i \frac {\nz^3}{16} \wunp^2 \wdos \wtresp - \frac {\nz^4}{32} \wunp^2 \wtresp^2
 , \nonumber \\
B_8 &=& \frac {\nz}{16} \wdos^2 -  \frac {\nz}{16} \wtres^2 - \frac {\nz}{4}
\wdosp^2 - \frac {\nz}{4} \wtresp^2 
- \frac {i}{2} \wdos \wtresp - \frac {i}{2} \wdosp \wtres 
+ \frac {i}{2} \wunp \wdos \wtres \nonumber \\&+&
\frac {\nz}{4} \wunp \wtres \wtresp - \frac {\nz}{4} \wunp \wdos \wdosp -
\frac {\nz}{16} \wunp^2 \wtres^2 + \frac {\nz}{16} \wunp^2 \wdos^2, 
\nonumber \\
B_9 &=& \frac {\nz}{12} + \frac {\nz}{12} \wunp^2  , \nonumber \\
B_{10} &=& - \frac {\nz}{12} - \frac {\nz}{12} \wunp^2  , \nonumber \\
B_{11} &=& - \frac {\nz}{24} + \frac {\nz}{24} \wunp^2  , \nonumber \\
B_{12} &=& \frac {\nz}{8} \wdos^2 + \frac {\nz}{8} \wtres^2 - 
\frac {\nz}{2} \wdosp^2 - \frac {\nz}{2} \wtresp^2 + i \wdosp \wtres -
 i \wdos \wtresp +\frac {\nz}{2} \wunp \wdos \wdosp  \nonumber \\
&+& \frac {\nz}{2} \wunp \wtres \wtresp - 
\frac {\nz}{8} \wunp^2 \wdos^2 - \frac {\nz}{8} \wunp^2 \wtres^2 , \nonumber \\
B_{13} &=& \frac {1}{4} , \nonumber \\
B_{14} &=& -\frac {\nz}{4} \wunpp + \frac {\nz}{2} \wunp^2 +
\frac {\nz}{3} \wunpp^2 +
\left( \frac {5 \nz}{12}- \frac {\nz^3}{8} \right) \wunp^2 \wunpp +
\left(- \frac {\nz}{8} + \frac {3 \nz^3}{16} \right)\wunp^4
 , \nonumber \\
B_{15} &=& - \frac {\nz}{3} \wunpp^2 + 
\frac {2 \nz}{3} \wunp^2 \wunpp - \frac {\nz}{4} \wunp^4 , \nonumber \\
B_{16} &=& - \frac {1}{4} \wunpp + \frac {3}{8} \wunp^2 + 
\left(\frac {\nz^2}{8} - \frac {1}{4}\right) \wunpp^2 +
\left(\frac {3}{4}-\frac {3\nz^2}{8}\right)\wunp^2 \wunpp +
\left( \frac {9\nz^2}{32}-\frac {9}{16} \right) \wunp^4 , \nonumber \\
B_{17} &=&  \frac {\nz}{8} \wunp \wdosp - \frac {\nz}{4} \wunpp \wdos + 
\frac {\nz}{4} \wunpp \wdospp - \frac {i}{8} \wunp \wtres +
\frac {\nz}{8} \wunp^2 \wdos - \frac {\nz^3}{8} \wunp \wunpp \wdosp 
\nonumber \\
&-& \frac {\nz}{4} \wunp^2\wdospp + \nonumber +
i\left(\frac {\nz^2}{8}-\frac {1}{4}  \right) \wunp \wunpp \wtres -
\frac {\nz}{8} \wunp^3 \wdos + \frac {3\nz^3}{16}\wunp^3 \wdosp 
\nonumber \\&+& 3i \left(\frac {1}{8}- \frac {\nz^2}{16} \right)\wunp^3 \wtres
 , \nonumber \\
B_{18} &=& -\frac {1}{4} \wdos - \frac {1}{2} \wunpp \wdos + 
i \frac {\nz}{2}\wunpp \wtresp + \frac {3}{4}\wunp^2\wdos-
i \frac {\nz}{4} \wunp \wunpp \wtres -  i \frac {\nz}{2} \wunp^2\wtresp 
 \nonumber \\ &+&i \frac {\nz}{4} \wunp^3 \wtres ,
\nonumber \\
B_{19} &=& \frac {1}{24} - \frac {1}{24}\wunp^2 , \nonumber \\
B_{20} &=& \frac {1}{12} + \frac {1}{12}\wunp^2 , \nonumber \\
B_{21} &=& - \frac {\nz}{8} \wtres + \frac {\nz}{8} \wunp^2 \wtres, 
\nonumber \\
B_{22} &=& -  \frac {1}{8}\wtres - i \frac {\nz}{8} \wunp \wdos -
i \frac {\nz}{2} \wunpp \wdosp - \frac {1}{4}\wunpp \wtres - 
\frac {\nz^2}{8}\wunp \wtresp-
\frac {\nz^2}{4}\wunpp \wtrespp  \nonumber \\ &+&
i \frac {\nz}{8}\wunp \wunpp \wdos +  i \frac {\nz}{2}\wunp^2 \wdosp +
\frac {1}{8} \left(3- \nz^2 \right)\wunp^2\wtres + 
\frac {\nz^2}{8}\wunp \wunpp \wtresp  \nonumber \\ &+& 
\frac {\nz^2}{4}\wunp^2\wtrespp - i \frac {\nz^3}{16} \wunp^3 \wdos - 
\frac {\nz^4}{16} \wunp^3 \wtresp, \nonumber \\
B_{23} &=& \frac {1}{4}\wtres +  \frac {1}{2} \wunpp \wtres +
i \frac {\nz}{2} \wunpp \wdosp - i \frac {\nz}{4}\wunp \wunpp \wdos +
i \frac {\nz}{2} \wunp^2 \wdosp 
- \frac {3}{4} \wunp^2 \wtres  \nonumber \\ &+&i \frac {\nz}{4} \wunp^3 \wdos ,
\nonumber \\
B_{24} &=& \frac {i}{8} \wunp \wdos - \frac {\nz}{4}\wunpp \wtres +
\frac {\nz}{8}\wunp \wtresp +  \frac {\nz}{4}\wunpp \wtrespp +
\frac {i}{4}\left(1- \frac {\nz^2}{2}\right)\wunp \wunpp \wdos 
\nonumber \\ &+&\frac {3\nz}{8}\wunp^2\wtres - \frac {\nz^3}{8}\wunp \wunpp 
\wtresp - \frac {\nz}{4} \wunp^2 \wtrespp +
\frac {3i}{8}\left(\frac {\nz^2}{2}-1\right)\wunp^3 \wdos 
\nonumber \\ &+&\left(- \frac {\nz}{8}+
\frac {3 \nz^3}{16}\right)\wunp^3 \wtresp, \nonumber \\
B_{25} &=& \frac {i}{2}\wdos\wdosp-\frac {i}{2}\wtres\wtresp +
\frac {\nz}{8}\wdos\wtres +\frac {\nz}{2}\wdosp\wtresp -
\frac {i}{4}\wunp \wdos^2 + \frac {i}{4}\wunp \wtres^2 \nonumber \\ &-&
\frac {\nz}{4}\wunp\wdosp\wtres -
\frac {\nz}{4}\wunp\wdos\wtresp +\frac {\nz}{8}\wunp^2\wdos\wtres 
, \nonumber \\
B_{26} &=& i \frac {\nz}{2}\wdosp\wdospp -i \frac {\nz}{2}\wtresp\wtrespp +
\frac {1}{8} \wdos\wtres +\frac {1}{2} \wdosp\wtresp +
 \frac {\nz^2}{4}\wdospp\wtrespp +
\frac {1}{4} \wdospp\wtres  \nonumber \\ &+&\frac {1}{4} \wdos\wtrespp +
i \frac {\nz}{8} \wunp \wdos^2
-i \frac {\nz}{4} \wunp \wdosp^2+
\frac{i \nz^3}{16} \wunp^2 \wdos \wdosp+
i \frac {\nz}{8} \wunp \wdos \wdospp  \nonumber \\ &+&
\frac {1}{4}\left(\frac {\nz^2}{2}-1\right)\wunp\wdosp\wtres -
i \frac {\nz}{8}\wunp\wtres^2 + 
\frac {1}{4}\left(\frac {\nz^2}{2}-1\right)\wunp\wdos\wtresp -
\frac {\nz^2}{8}\wunp\wdospp\wtresp  \nonumber \\ &+& 
i \frac {\nz}{4}\wunp\wtresp^2 -
\frac {\nz^2}{8}\wunp\wdosp\wtrespp +i \frac {\nz}{8}\wunp\wtres\wtrespp +
\frac {1}{8}\left(\frac {\nz^2}{2}-1\right)\wunp^2\wdos\wtres 
 \nonumber \\ &+&\frac {\nz^4}{16} \wunp^2\wdosp\wtresp - i \frac {\nz^3}{16}
\wunp^2\wtres\wtresp, \nonumber \\
B_{27} &=& -i\frac {\nz}{6} \wunp \wunpp + i\frac {\nz}{6 } \wunp^3
, \nonumber \\
B_{28} &=& 0, \nonumber \\
B_{29} &=& 0, \nonumber \\
B_{30} &=& 0, \nonumber \\
B_{31} &=& 0, \nonumber \\
B_{32} &=& \frac {\nz}{2}\wunpp - \frac {\nz}{2}\wunp^2 - 
\frac {\nz}{6}\wunpp^2 + \left(-\frac {\nz}{6} +
\frac {\nz^3}{4} \right)\wunp^2 \wunpp +
\left(\frac {\nz}{3} - \frac {\nz^3}{4}\right)\wunp^4, \nonumber \\
B_{33} &=& \frac {2\nz}{3}\wunpp^2 -
\frac {4 \nz}{3} \wunp^2\wunpp +
\frac {2 \nz}{3} \wunp^4, \nonumber \\
B_{34} &=& -\frac {1}{2}\wunpp +\frac {1}{2}\wunp^2 +
\left(\frac {\nz^2}{2}-1\right) \wunpp^2 +
\frac {5}{2}\left(1-\frac {\nz^2}{2}\right) \wunp^2\wunpp -
\frac {3}{2}\left(1-\frac {\nz^2}{2}\right)  \wunp^4, \nonumber \\
B_{35} &=& \frac {\nz}{4}\wunpp - \frac {\nz}{8}\wunp^2 +
\frac {\nz}{6}\wunpp^2 + 
 \frac {\nz^2}{8}\wunp\wcincp +  \frac {\nz^2}{4}\wunpp\wcincpp - 
\frac {\nz^2}{8}\wunp\wsisp  \nonumber \\ &-&\frac {\nz^2}{4}\wunpp\wsispp +
\left( \frac {11 \nz}{24}+ \frac {\nz^3}{8}\right) \wunp^2\wunpp -
 \frac {\nz^2}{8}\wunp\wunpp\wcincp -  \frac {\nz^2}{4}\wunp^2\wcincpp 
\nonumber \\ &+&\frac {\nz^2}{8}\wunp\wunpp\wsisp +
 \frac {\nz^2}{4}\wunp^2\wsispp +
\left(\frac {\nz}{6} - \frac {\nz^3}{16}\right) \wunp^4 +
 \frac {\nz^4}{16} \wunp^3 \wcincp \nonumber \\ &-&
\frac {\nz^4}{16} \wunp^3 \wsisp, \nonumber \\
B_{36} &=& \frac {\nz}{3}\wunpp^2-\frac {2\nz}{3}\wunp^2\wunpp +
\frac {\nz}{3}\wunp^4, \nonumber \\
B_{37} &=& -\frac {1}{4}\wunpp + \frac {1}{8}\wunp^2 +
\left(\frac {\nz^2}{4}-\frac {5}{12}\right)\wunpp^2 -
 \frac {\nz}{8}\wunp\wcincp -  \frac {\nz}{4}\wunpp\wcincpp +
\frac {\nz}{8}\wunp\wsisp \nonumber \\ &+& \frac {\nz}{4}\wunpp\wsispp +
\left(\frac {5}{6}-\frac {\nz^2}{2}\right)\wunp^2\wunpp +
 \frac {\nz^3}{8}\wunp\wunpp\wcincp - \frac {\nz^3}{8}\wunp\wunpp\wsisp 
\nonumber \\ &+& \frac {1}{4}\wunp^2\wcincpp -
\frac {\nz}{4}\wunp^2\wsispp +
\left(\frac {3\nz^2}{16}-\frac {7}{24}\right)\wunp^4 +
 \left(\frac {\nz}{8}-\frac {3 \nz^3}{16}\right)\wunp^3\wcincp 
\nonumber \\ &+&\left(- \frac {\nz}{8}+
\frac {3 \nz^3}{16}\right)\wunp^3\wsisp, \nonumber \\
B_{38} &=& \left(\frac {\nz^2}{2}-\frac {5}{6}\right)\wunpp^2 +
\left(\frac {5}{3}-\nz^2\right)\wunp^2\wunpp +
\left(\frac {\nz^2}{2}-\frac {5}{6}\right)\wunp^4, \nonumber \\
B_{39} &=& \frac {1}{2}\left(\nz^2-1\right)\wunpp^2 -
\frac {3}{4}\left(\nz^2-1\right)\wunp^2\wunpp +
\frac {1}{4}\left(\nz^3-\nz\right)\wunp\wunpp\wcincp \nonumber \\ &-&
\frac {1}{4}\left(\nz^3-\nz\right)\wunp\wunpp\wsisp +
\frac {1}{4}\left(\nz^2-1\right)\wunp^4 -
\frac {1}{4}\left(\nz^3-\nz\right)\wunp^3\wcincp \nonumber \\ &+&
\frac {1}{4}\left(\nz^3-\nz\right)\wunp^3\wsisp , \nonumber \\
B_{40} &=&\frac {\nz^2}{8}\wunpp^2 - \frac {\nz^2}{4}\wcincpp\wsispp +
 \frac {\nz}{4}\wunpp\wcincpp - \frac {\nz}{4}\wunpp\wsispp + 
\frac {\nz^2}{8}\wcincpp^2 +\frac {\nz^2}{8}\wsispp^2 
\nonumber \\ &-&\frac {\nz^2}{8}\wunp^2\wunpp +
\frac {1}{4}\left(\frac {\nz^3}{2}-\nz\right)\wunp\wunpp\wcincp -
\frac {1}{4}\left(\frac {\nz^3}{2}-\nz\right)\wunp\wunpp\wsisp 
\nonumber \\ &-& \frac {\nz}{8}\wunp^2\wcincp +\frac {\nz}{8}\wunp^2\wsisp
-
\frac {\nz^2}{8}\wunp\wcincp\wcincpp - \frac {\nz^2}{8}\wunp\wsisp\wsispp +
 \frac {\nz^2}{8}\wunp\wcincpp\wsisp \nonumber \\ &+&
 \frac {\nz^2}{8}\wunp\wcincp\wsispp +
\frac {\nz^2}{32}\wunp^4 -
\frac {1}{8}\left(\frac {\nz^3}{2}-\nz\right)\wunp^3\wcincp +
\frac {1}{8}\left(\frac {\nz^3}{2}-\nz\right)\wunp^3\wsisp 
\nonumber \\ &+&\frac {\nz^4}{32}\wunp^2\wcincp^2 + 
\frac {\nz^4}{32}\wunp^2\wsisp^2 -
 \frac {\nz^4}{16}\wunp^2\wcincp\wsisp , \nonumber \\
B_{41} &=& -i \frac {\nz}{4}\wunp -i \frac {\nz}{4}\wunp\wunpp -i
\frac {\nz^2}{8}\wunp\wcinc -i\frac {\nz^2}{4}\wunpp\wcincp +
i\left(\frac {\nz}{2} - \frac {\nz^3}{8}\right)\wunp^3 
\nonumber \\ &+&i\frac {\nz^2}{8}\wunp\wunpp\wcinc +i 
\frac {\nz^2}{4}\wunp^2\wcincp -i
\frac {\nz^4}{16}\wunp^3\wcinc, \nonumber \\
B_{42} &=&\frac {i}{4}\wunp +
\frac {i}{2}\left(1-\frac {\nz^2}{2}\right)\wunp\wunpp +i
\frac {\nz}{8}\wunp\wcinc +i \frac {\nz}{4}\wunpp\wcincp +
i \left(- \frac {3}{4} + \frac {3 \nz^3}{16} \right)\wunp^3 
\nonumber \\ &-&i\frac {\nz^3}{8}\wunp\wunpp\wcinc -i
\frac {\nz}{4}\wunp^2\wcincp -i
\left(\frac {\nz}{8} - \frac {3 \nz^3}{16}\right)\wunp^3\wcinc , \nonumber \\
B_{43} &=&\frac {i}{2}\left(1-\nz^2\right)\wunp\wunpp -
\frac {i}{2}\left(1-\nz^2\right)\wunp^3 -
\frac {i}{4}\left(\nz^3-\nz\right)\wunp\wunpp\wcinc \nonumber \\ &+&
\frac {i}{4}\left(\nz^3-\nz\right)\wunp^3\wcinc, \nonumber \\
B_{44} &=& -i \frac {\nz^2}{4}\wunp\wunpp -i \frac {\nz}{4}\wunpp\wcincp
-i
\frac {\nz}{4}\wunp\wcincpp - i \frac {\nz^2}{4}\wcincp\wcincpp +
i \frac {\nz}{4}\wunp\wsispp \nonumber \\ &+&i\frac {\nz^2}{4}\wcincp\wsispp +
i \frac {\nz^2}{8}\wunp^3 -
\frac {i}{4}\left(\frac {\nz^3}{2}-\nz\right)\wunp\wunpp\wcinc -
\frac {i}{8}\left(\nz^3-3\nz\right)\wunp^2\wcincp 
\nonumber \\ &+&i \frac {\nz^2}{8}\wunp\wcincp^2 + 
 i \frac {\nz^2}{8}\wunp\wcinc\wcincpp +
\frac {i}{4}\left(\frac {\nz^3}{2}-\nz\right)\wunp^2\wsisp -i
\frac {\nz^2}{8}\wunp\wcincp\wsisp \nonumber \\ &-&i
\frac {\nz^2}{8}\wunp\wcinc\wsispp +
\frac {i}{8}\left(\frac {\nz^3}{2}-\nz\right)\wunp^3\wcinc -
i \frac {\nz^4}{16}\wunp^2\wcinc\wcincp +i 
\frac {\nz^4}{16}\wunp^2\wcinc\wsisp, \nonumber \\
B_{45} &=&\frac {\nz^2}{8}\wunp^2 +  \frac {\nz}{4}\wunp\wcincp +
\frac {\nz^2}{8}\wcincp^2 +
\frac {1}{4}\left(\frac {\nz^3}{2}-\nz\right)\wunp^2\wcinc -
\frac {\nz^2}{8}\wunp\wcinc\wcincp  \nonumber \\ &+&
\frac {\nz^4}{32}\wunp^2\wcinc^2, \nonumber \\
B_{46} &=& \frac {1}{2}\wunpp\wdos - i \frac {\nz}{2}\wunpp\wtresp -
\frac {1}{2}\wunp^2\wdos+ i \frac {\nz}{4}\wunp\wunpp\wtres + 
i \frac {\nz}{2}\wunp^2\wtresp - i \frac {\nz}{4}\wunp^3\wtres , \nonumber \\
B_{47} &=& -\frac {\nz}{2}\wunpp\wdos + \frac {i}{2}\wunpp\wtresp +
\frac {\nz}{2}\wunp^2\wdos + 
\frac {1}{4}\left(\nz-\nz^3\right)\wunp\wunpp\wdosp  \nonumber\\
&+& i \left( - \frac {1}{2} + \frac {\nz^2}{4}\right)\wunp\wunpp\wtres 
-\frac {i}{2}\wunp^2\wtresp -
\frac {1}{4}\left(\nz-\nz^3\right)\wunp^3\wdosp +
\frac {i}{2}\left(1-\nz^2\right)\wunp^3\wtres, \nonumber \\
B_{48} &=& i \frac {\nz}{2}\wunpp\wdosp + \frac {i}{2}\wunpp\wtres -
i \frac {\nz}{4}\wunp\wunpp\wdos - i \frac {\nz}{2}\wunp^2\wdosp -
\frac {1}{2}\wunp^2\wtres + i \frac {\nz}{4}\wunp^3\wdos, \nonumber \\
B_{49} &=&- \frac {i}{2}\wunpp\wdosp - \frac {\nz}{2}\wunpp\wtres +
\frac {i}{2}\left(1-\frac {\nz^2}{2}\right)\wunp\wunpp\wdos +
\frac {i}{2}\wunp^2\wdosp + \frac {\nz}{2}\wunp^2\wtres \nonumber \\ &+&
\frac {1}{4}\left(\nz-\nz^3\right)\wunp\wunpp\wtresp -
\frac {i}{2}\left(1-\frac {\nz^2}{2}\right)\wunp^3\wdos -
\frac {1}{4}\left(\nz-\nz^3\right)\wunp^3\wtresp, \nonumber \\
B_{50} &=&-\frac {\nz}{4}\wunpp\wdos 
- \frac {\nz}{4}\wunpp\wdospp 
+\frac {i}{2}\wunpp\wtresp 
-\frac {1}{4}\wcincpp\wdos 
- \frac {\nz^2}{4}\wcincpp\wdospp  
+\frac {1}{4}\wsispp\wdos 
\nonumber \\ &+&\frac {\nz^2}{4}\wsispp\wdospp 
+i\frac {\nz}{2}\wcincpp\wtresp 
-i \frac {\nz}{2}\wsispp\wtresp 
+\frac {\nz}{8}\wunp^2\wdos 
+ \frac {\nz}{8}\wunp^2\wdospp 
\nonumber \\ &-&\frac {1}{4}\left(\frac {\nz^3}{2}-\nz\right)\wunp\wunpp\wdosp 
+\frac {i}{4}\left(\frac {\nz^2}{2}-1\right)\wunp\wunpp\wtres 
-\frac {i}{4}\wunp^2\wtresp 
\nonumber \\ &-&\frac {1}{4}\left(\frac {\nz^2}{2}-1\right)
\wunp\wdos\wcincp 
+ \frac {\nz^2}{8}\wunp\wdospp\wcincp 
-i\frac {\nz}{4}\wunp\wtresp\wcincp 
+ \frac {\nz^2}{8}\wunp\wdosp\wcincpp
\nonumber \\ &-&i\frac {\nz}{8}\wunp\wtres\wcincpp 
+\frac {1}{4}\left(\frac {\nz^2}{2}-1\right)\wunp\wdos\wsisp 
-\frac {\nz^2}{8}\wunp\wdospp\wsisp 
+i \frac {\nz}{4}\wunp\wtresp\wsisp 
\nonumber \\ &-&\frac {\nz^2}{8}\wunp\wdosp\wsispp
+i \frac {\nz}{8}\wunp\wtres\wsispp 
+\frac {1}{8}\left(\frac {\nz^3}{2}-\nz\right)\wunp^3\wdosp 
-\frac {i}{8}\left(\frac {\nz^2}{2}-1\right)\wunp^3\wtres 
\nonumber \\ &-& \frac {\nz^4}{16}\wunp^2\wdosp\wcincp 
+i\frac {\nz^3}{16}\wunp^2\wtres\wcincp 
+\frac {\nz^4}{16}\wunp^2\wdosp\wsisp 
-i \frac {\nz^3}{16}\wunp^2\wtres\wsisp 
, \nonumber \\ 
B_{51} &=&  - \frac {\nz}{4}\wunpp\wtres
- \frac {\nz}{4}\wunpp\wtrespp
-\frac{i}{2}\wunpp\wdosp
- \frac {1}{4}\wcincpp\wtres
- \frac {\nz^2}{4}\wcincpp\wtrespp 
+ \frac {1}{4}\wsispp\wtres 
\nonumber \\ &+&\frac {\nz^2}{4}\wsispp\wtrespp 
-i \frac {\nz}{2}\wcincpp\wdosp
+ i \frac {\nz}{2}\wsispp\wdosp
+ \frac {\nz}{8}\wunp^2\wtres
+\frac {\nz}{8}\wunp^2\wtrespp
\nonumber \\ &-&\frac {1}{4}\left(\frac {\nz^3}{2}-\nz\right)
\wunp\wunpp\wtresp 
-\frac {i}{4}\left(\frac {\nz^2}{2}-1\right)\wunp\wunpp\wdos
+\frac {i}{4} \wunp^2\wdosp 
\nonumber \\ &-&\frac {1}{4}\left(\frac {\nz^2}{2}-1\right)
\wunp\wtres\wcincp
+ \frac {\nz^2}{8}\wunp\wtrespp\wcincp
+i\frac {\nz}{4}\wunp\wdosp\wcincp
+ \frac {\nz^2}{8}\wunp\wtresp\wcincpp
\nonumber \\ &+&i\frac {\nz}{8}\wunp\wdos\wcincpp
+\frac {1}{4}\left(\frac {\nz^2}{2}-1\right)\wunp\wtres\wsisp 
-\frac {\nz^2}{8}\wunp\wtrespp\wsisp
-i \frac {\nz}{4}\wunp\wdosp\wsisp 
\nonumber \\ &-&\frac {\nz^2}{8}\wunp\wtresp\wsispp 
- i \frac {\nz}{8}\wunp\wdos\wsispp 
+\frac {1}{8}\left(\frac {\nz^3}{2}-\nz\right)\wunp^3\wtresp 
+\frac {i}{8}\left(\frac {\nz^2}{2}-1\right)\wunp^3\wdos 
\nonumber \\ &-& \frac {\nz^4}{16}\wunp^2\wtresp\wcincp 
-i \frac {\nz^3}{16}\wunp^2\wdos\wcincp 
+\frac {\nz^4}{16}\wunp^2\wtresp\wsisp 
+i \frac {\nz^3}{16}\wunp^2\wdos\wsisp 
, \nonumber \\ 
B_{52} &=& i \frac {\nz}{4}\wunp\wdos +i \frac {\nz}{4}\wunp\wdospp +
\frac {1}{2}\wunp\wtresp + \frac {i}{4}\wdos\wcincp +i 
\frac {\nz^2}{4}\wdospp\wcincp  +  \frac {\nz}{2}\wtresp\wcincp 
\nonumber \\ &+&
\frac {i}{4}\left(\frac {\nz^3}{2}-\nz\right)\wunp^2\wdosp +
\frac {1}{4}\left(\frac {\nz^2}{2}-1\right)\wunp^2\wtres +
\frac {i}{4}\left(\frac {\nz^2}{2}-1\right)\wunp\wdos\wcinc 
\nonumber \\ &-&i\frac {\nz^2}{8}\wunp\wdospp\wcinc -
 \frac {\nz}{4}\wunp\wtresp\wcinc -i  
\frac {\nz^2}{8}\wunp\wdosp\wcincp -  \frac {\nz}{8}\wunp\wtres\wcincp 
\nonumber \\ &+&i\frac {\nz^4}{16}\wunp^2\wdosp\wcinc +
 \frac {\nz^3}{16}\wunp^2\wtres\wcinc 
, \nonumber \\ 
B_{53} &=&  -i \frac {\nz}{4}\wunp\wtres -i \frac {\nz}{4}\wunp\wtrespp 
+\frac {1}{2}\wunp\wdosp -\frac {i}{4}\wtres\wcincp  -
i\frac {\nz^2}{4}\wtrespp\wcincp +  \frac {\nz}{2}\wdosp\wcincp \nonumber \\ 
&+& \frac {1}{4}\left(\frac {\nz^2}{2}-1\right)\wunp^2\wdos -
\frac {i}{4}\left(\frac {\nz^3}{2}-\nz\right)\wunp^2\wtresp -
\frac {i}{4}\left(\frac {\nz^2}{2}-1\right)\wunp\wtres\wcinc 
\nonumber \\ &+&i\frac {\nz^2}{8}\wunp\wtrespp\wcinc - 
 \frac {\nz}{4}\wunp\wdosp\wcinc + i
\frac {\nz^2}{8}\wunp\wtresp\wcincp -  \frac {\nz}{8}\wunp\wdos\wcincp 
\nonumber \\ &-&i\frac {\nz^4}{16}\wunp^2\wtresp\wcinc +
 \frac {\nz^3}{16}\wunp^2\wdos\wcinc, \nonumber \\ 
B_{54} &=& i \frac {\nz}{6}\wunp^3 -i \frac {\nz}{6}\wunp\wunpp, \nonumber \\
B_{55} &=& -\frac {i}{6}\wunp^3 +\frac {i}{6}\wunp\wunpp, \nonumber \\
B_{56} &=& 0. \nonumber \\
B_{57} &=& 0. \nonumber \\
\end{eqnarray}

\section{Appendix E: The Discrete Symmetries: C, P and T.}

For the sake of completeness we give in this appendix the transformation
laws for the fields and the sources that appear in the article under
the discrete symmetries, C, P and T.
In order to specify them one needs to first define how they act on the 
space-time coordinates $x^\mu=(t, \vec{x})$.

The Parity (P) operation transforms $\vec{x} \to -\vec{x}$ while leaving
the time component unchanged. 
Using the Minkowski space notation, $x^\mu \buildrel P \over \rightarrow
x_p^\mu= p^\mu_{\; \nu} x^\nu$,
where $p^\mu_{\; \nu}= \ diagonal \ (1, -1,-1,-1)$ is a matrix.

Time-Reversal (T) reverses the flow of the time-component
$t \to -t$ while leaving the space components unchanged.
$x^\mu \buildrel T \over \rightarrow x_t^\mu= t^\mu_{\;\nu} x^\nu$,
where $t^\mu_{\;\nu}= \ diagonal \ (-1,1,1,1)$.

Charge-Conjugation (C) does not act on space time-indices, it
interchanges the r\^ole of particles and anti-particles.

In Quantum Mechanics they are implemented with operators acting
on a Hilbert space that are unitary for C, P; and anti-unitary 
for T.

Acting on the (Dirac) quark fields $q_a(x)$, where $a$ labels
any colour or flavour index, they read
\begin{eqnarray}
q_a(x)  \buildrel C \over \longrightarrow q_a^{(C)}(x) &=& \xi_C
{\cal C}   {\bar{q_a}}^T (x), \;\;\;\;\;\;
{\cal C} \gamma_\mu^T {\cal C}^{-1}=
- \gamma_\mu;  \nonumber\\
\buildrel P \over \longrightarrow q_a^{(P)}(x) &=& \xi_P
{\cal P}   q_a (x_p), \;\;\;\;\;\;
{\cal P} \gamma_\mu^\dagger {\cal P}^{-1}=\gamma_\mu; \nonumber\\
\buildrel T \over \longrightarrow q_a^{(T)}(x) &=& \xi_T
{\cal T}  q_a (x_t), \;\;\;\;\;\; 
{\cal T} \gamma_\mu^T {\cal T}^{-1}= \gamma_\mu, 
\label{quarks}
\end{eqnarray}
The $\xi_C$, $\xi_P$, $\xi_T$ are arbitrary phase factors,  
$|\xi_C|^2=|\xi_P|^2=|\xi_T|^2=1$.
The matrices ${\cal C}$, ${\cal P}$, ${\cal T}$ act on Dirac indices
only. In the Dirac representation ${\cal C}= i \gamma^0 \gamma^2$,
${\cal P}= \gamma^0$. Once ${\cal C}$, ${\cal P}$ are fixed, ${\cal T}=
-i \gamma_5 {\cal C}$. They verify 
${\cal C}^{-1}={\cal C}^\dagger={\cal C}^T=-{\cal C}$, and 
${\cal T}^{-1}={\cal T}^\dagger=-{\cal T}^T=
{\cal T}$.
Acting on $\gamma_5$ they yield
$${\cal C} \gamma_5^T {\cal C}^{-1}= \gamma_5, \;\;\;\;
\gamma^0 \gamma_5  \gamma^0= -\gamma_5, \;\;\;\;
{\cal T} \gamma_5^T {\cal T}= \gamma_5.$$

For $\bar{q}_a (x)$,
\begin{eqnarray}
\bar{q}_a(x) & \buildrel C \over \longrightarrow & 
- \xi_C^* q_a^T(x) {\cal C}^{-1},
\nonumber\\
&\buildrel P \over \longrightarrow &  \;\;\; \xi_P^*
\bar{q}_a (x_p) \gamma_0,
\nonumber\\  
&\buildrel T \over \longrightarrow & \;\;\; \xi_T^*
\bar{q_a}(x_t) {\cal T}, 
\label{antiquarks}
\end{eqnarray}
The phase factors $\xi_C$, $\xi_P$, $\xi_T$ shall be omitted henceforth.
The quark bilinears $\bar{q}_a (x) \Gamma q_b (x)$ transform as
\begin{eqnarray}
&\buildrel C \over \longrightarrow& \bar{q}_b (x) [\Gamma]_C \; q_a (x),
\;\;\;\;\;\;\;\;\; [\Gamma]_C \;=
\left( {\cal C}^{-1} \Gamma {\cal C} \right)^T,
\nonumber\\
&\buildrel P \over \longrightarrow& \bar{q}_a (x_p) [\Gamma]_P \; q_b (x_p),
\;\;\;\;\;\;[\Gamma]_P \; = \gamma^0 \Gamma \gamma^0,
\nonumber\\
&\buildrel T \over \longrightarrow& \bar{q}_a (x_t) [\Gamma]_T \; q_b (x_t),
\;\;\;\;\;\;\;[\Gamma]_T \; = {\cal T} \Gamma^* {\cal T}. 
\label{bilinears}
\end{eqnarray}
The star $(*)$ in the last line of (\ref{bilinears}) denotes complex
conjugation. There is a minus sign in the bilinear transformed under C
which comes from the anti-commutation of two quark fields.
$${[I]}_C= I, \;\;\;\;\;\;\;\;\;\; {[I]}_P= I ,
\;\;\;\;\;\;\;\;\;\; {[I]}_T= I, $$
$$[i {\gamma}_5]_C=i \gamma_5, \;\;\;\;\;\;\; 
[i {\gamma}_5]_P= - i {\gamma}_5, 
\;\;\;\;\;\;\;
[i {\gamma}_5]_T= - i {\gamma}_5,$$
$$[\gamma^\mu]_C = - \gamma^\mu, \;\;\;\;\;\; [\gamma^\mu]_P= p^\mu_{\;\mu'} 
[\gamma^{\mu'}],
\;\;\;\;\;\;
[\gamma^\mu]_T= -t^\mu_{\;\mu'} [\gamma^{\mu'}], $$
$$[\gamma^\mu \gamma_5]_C= [\gamma^\mu \gamma_5], \;\; 
[\gamma^\mu \gamma_5]_P= -p^\mu_{\;\mu'}[\gamma^{\mu'} \gamma_5], \;\;
[\gamma^\mu \gamma_5]_T= -t^\mu_{\;\mu'}[\gamma^{\mu'} \gamma_5].$$
  
For the gluon field (hermitian) matrix $G^\mu (x)$ in colour-space,

$$\buildrel C \over \rightarrow - G^{\mu \; T} (x),
\;\;\;\; \buildrel P \over \rightarrow p^\mu_{\;\mu'} G^{\mu'} (x_p),
\;\;\;\; \buildrel T \over \rightarrow -t^\mu_{\;\mu'} G^{\mu'} (x_t).$$

It is easy to verify that the QCD action is invariant under C, P and T.

For the topological charge density $Q(x) \sim \epsilon_{\mu \nu \rho \sigma}
\; Tr_c G^{\mu \nu}(x) G^{\rho \sigma}(x)$, which is also real,
$$\buildrel C \over \rightarrow  Q(x),
\;\;\;\; \buildrel P \over \rightarrow \det (p^\mu_{\;\mu'}) \ Q(x_p)= -Q(x_p),
\;\;\;\; \buildrel T \over \rightarrow \det (t^\mu_{\;\mu'}) \ Q(x_t)= -Q(x_t).$$
So far for operators involving the dynamical fields.

The (hermitian) operator $i \bar{q}_a (x)\gamma_5 q_b(x)$ 
has the same quantum 
as the light pseudoscalar matrix $\Phi_{ab}(x)$, and,
since the vacuum is invariant under C, P, and T the transformation
laws of the latter are taken from those of the former.
Let us write them down for the simpler case of $U_L(2) \otimes U_R(2)$, where
\begin{equation}
\Phi =
\left(\begin{array}{cc} 
\frac {\pi_0 - \eta}{\sqrt{2}} &  \pi^+ \\
\pi^- & -\frac {\pi_0 + \eta} {\sqrt{2}}\\
\end{array}\right).
\label{pions}
\end{equation}
Under the discrete symmetries
\begin{equation}
\buildrel C \over \longrightarrow 
\left(\begin{array}{cc} 
\frac{\pi_0 - \eta}{\sqrt{2}}  &   \pi^-  \\
\pi^+ & - \frac {\pi_0 (x) + \eta}{\sqrt{2}}\\
\end{array}\right) = \Phi^T,  \nonumber
\end{equation}
and, similarly,
\begin{equation}
\begin{array}{cc}
\buildrel P  \over \longrightarrow - \Phi (x_p), & 
\buildrel T \over \longrightarrow  - \Phi (x_t),
\end{array}
\end{equation}
which generalize immediately for $U_L (n_l) \otimes U_R (n_l)$. It translates
into
\begin{eqnarray}
U(x) &\buildrel \rm{C} \over \longrightarrow& U^{(C)}(x)=U^T (x), 
\nonumber \\
U(x) &\buildrel \rm{P} \over \longrightarrow& U^{(P)}(x)=U^\dagger (x_p),
\nonumber \\
U(x) &\buildrel \rm{T} \over \longrightarrow& U^{(T)}(x)=U(x_t),
\label{cptu}
\end{eqnarray}
for the $U(x)$ matrix, as defined in (\ref{U}).

As for the external sources, we shall chose their transformation so as
to leave the action invariant.
The real source $\theta(x)$, that couples to the topogical charge $Q(x)$, transforms as:
$$\buildrel C \over \rightarrow  \theta(x),
\;\;\;\; \buildrel P \over \rightarrow -\theta (x_p),
\;\;\;\; \buildrel T \over \rightarrow -\theta (x_t).$$
In the text, the combination $\hat{\theta}= i \theta$ appeared in a natural
way. It transforms accordingly, with an extra minus sign for the
T transformation because it anti-commutes with the imaginary number $i$
due to its anti-unitary character. The combination $X$, defined in
(\ref{3.5bis}), transforms as $\hat{\theta}$ does.

The (hermitian) source matrices $s\ , p\ , v_\mu \ ,  a_\mu$, in flavour space
transform as
$$ s(x) \buildrel C \over \rightarrow  s^T(x),
\;\;\;\; \buildrel P \over \rightarrow  s(x_p),
\;\;\;\; \buildrel T \over \rightarrow  s(x_t),$$
$$ p(x) \buildrel C \over \rightarrow  p^T(x),
\;\;\;\; \buildrel P \over \rightarrow -p(x_p),
\;\;\;\; \buildrel T \over \rightarrow -p (x_t),$$
$$ v^\mu (x)\buildrel C \over \rightarrow - v^{\mu \; T}(x),
\;\;\;\; \buildrel P \over \rightarrow p^\mu_{\;\mu'}v^{\mu'}(x_p),
\;\;\;\; \buildrel T \over \rightarrow -t^\mu_{\;\mu'}v^{\mu'}(x_t),$$
$$ a^\mu (x)\buildrel C \over \rightarrow a^{\mu \; T}(x),
\;\;\;\; \buildrel P \over \rightarrow -p^\mu_{\;\mu'}a^{\mu'}(x_p),
\;\;\;\; \buildrel T \over \rightarrow -t^\mu_{\;\mu'}a^{\mu'}(x_t).$$

The combination $\chi=2 B (s +i p)$ transforms as the $U$ fields.

The left and right combinations of the vector and axial sources,
$l_\mu= v_\mu - a_\mu$ and $r_\mu= v_\mu + a_\mu$, transform as
$$ l^\mu (x)\buildrel C \over \rightarrow - r^{\mu \; T}(x),
\;\;\;\; \buildrel P \over \rightarrow p^\mu_{\;\mu'} r^{\mu'}(x_p),
\;\;\;\; \buildrel T \over \rightarrow -t^\mu_{\;\mu'} l^{\mu'}(x_t),$$
$$ r^\mu (x)\buildrel C \over \rightarrow -l^{\mu \; T}(x),
\;\;\;\; \buildrel P \over \rightarrow p^\mu_{\;\mu'} l^{\mu'}(x_p),
\;\;\;\; \buildrel T \over \rightarrow -t^\mu_{\;\mu'} r^{\mu'}(x_t).$$
Both the C and the P transformations interchange {\it left} and {\it right}.

For the field  strengths $F_L^{\mu \nu}$, $F_R^{\mu \nu}$ associated to
$l_\mu$, $r_\mu$,
$$F_L^{\mu \nu}(x)\buildrel C \over \rightarrow - F_R^{\mu \nu \; T}(x),
\;\;\;\; \buildrel P \over \rightarrow p^\mu_{\;\mu'} p^\nu_{\;\nu'}
F_R^{\mu' \nu'} (x_p),
\;\;\;\; \buildrel T \over \rightarrow -t^\mu_{\;\mu'} t^\nu_{\;\nu'}
F_L^{\mu' \nu'}(x_t),$$
$$F_R^{\mu \nu}(x)\buildrel C \over \rightarrow - F_L^{\mu \nu \; T}(x),
\;\;\;\; \buildrel P \over \rightarrow p^\mu_{\;\mu'} p^\nu_{\;\nu'}
F_L^{\mu' \nu'} (x_p),
\;\;\;\; \buildrel T \over \rightarrow -t^\mu_{\;\mu'} t^\nu_{\;\nu'}
F_R^{\mu' \nu'}(x_t),$$

Finally, for the combination $C^\mu= U^\dagger D^\mu U$, that is anti-hermitian
$C^{\mu \; \dagger}= - C^\mu$,
$$ C^\mu (x)\buildrel C \over \rightarrow [U(x) C^\mu (x) U^\dagger (x)]^T,
\;\;\;\; \buildrel P \over \rightarrow - p^\mu_{\;\mu'}
U(x_p) C^{\mu'} (x_p) U^\dagger (x_p),
\;\;\;\; \buildrel T \over \rightarrow t^\mu_{\;\mu'} C^{\mu'} (x_t).$$

\end{document}